\documentclass[numberedappendix,twocolappendix,iop,apj]{emulateapj}
\usepackage{psfig,amsfonts,amsmath,graphicx,natbib,nicefrac,url,hyperref,breakurl,longtable,graphicx
}

\usepackage{booktabs}
\hypersetup{colorlinks=true, urlcolor=blue, citecolor=blue}

\newcommand{\Swift}{\textit{Swift}}

\newcommand{\Chandra}{\textit{Chandra}}

\newcommand{\EKiso}{\ensuremath{E_{\rm K,iso}}}

\newcommand{\etarad}{\ensuremath{\eta_{\rm rad}}}
\newcommand{\epse}{\ensuremath{\epsilon_{\rm e}}}
\newcommand{\epsb}{\ensuremath{\epsilon_{\rm B}}}
\newcommand{\dens}{\ensuremath{n_{0}}}
\newcommand{\Astar}{\ensuremath{A_{*}}}
\newcommand{\tdec}{\ensuremath{t_{\rm dec}}}
\newcommand{\tjet}{\ensuremath{t_{\rm jet}}}
\newcommand{\thetajet}{\ensuremath{\theta_{\rm jet}}}
\newcommand{\tnr}{\ensuremath{t_{\rm NR}}}
\newcommand{\AV}{\ensuremath{A_{\rm V}}}

\newcommand{\pcmsq}{\ensuremath{{\rm cm}^{-2}}}
\newcommand{\pcc}{\ensuremath{{\rm cm}^{-3}}}

\newcommand{\Gammajet}{\ensuremath{\Gamma_{0}}}
\newcommand{\lsedov}{\ensuremath{l_{\rm Sedov}}}

\newcommand{\nua}{\ensuremath{\nu_{\rm a}}}

\newcommand{\numax}{\ensuremath{\nu_{\rm m}}}
\newcommand{\nuc}{\ensuremath{\nu_{\rm c}}}
\newcommand{\nuaf}{\ensuremath{\nu_{\rm a,f}}}
\newcommand{\numf}{\ensuremath{\nu_{\rm m,f}}}
\newcommand{\nucf}{\ensuremath{\nu_{\rm c,f}}}
\newcommand{\fnumf}{\ensuremath{F_{\nu,\rm m,f}}}
\newcommand{\nuar}{\ensuremath{\nu_{\rm a,r}}}
\newcommand{\numr}{\ensuremath{\nu_{\rm m,r}}}
\newcommand{\nucr}{\ensuremath{\nu_{\rm c,r}}}
\newcommand{\fnumr}{\ensuremath{F_{\nu,\rm m,r}}}

\newcommand{\RB}{\ensuremath{R_{\rm B}}}

\newcommand{\nuopt}{\ensuremath{\nu_{\rm opt}}}
\newcommand{\nux}{\ensuremath{\nu_{\rm X}}}

\newcommand{\fnumax}{\ensuremath{F_{\nu,\rm m}}}
\newcommand{\td}{\ensuremath{t_{\rm d}}}

\shorttitle{GRB~181201A}
\shortauthors{Laskar et al.}

\def\bath{1}
\def\northwestern{2}
\def\cierafellow{*}
\def\cssac{3}
\def\cfa{4}
\def\mpe{5}
\def\ou{6}
\def\ngslov{7}
\def\lvjm{8}
\def\mpir{9}
\def\huji{10}
\def\aoyama{11}
\def\mossuphys{12}
\def\mossuastro{13}
\def\oafa{14}
\def\icate{15}
\def\saao{16}
\def\pulkovo{17}
\def\iac{18}

\begin{document} 
\submitted{Published in ApJ}

\title{A Reverse Shock in GRB~181201A}
\author{Tanmoy~Laskar\altaffilmark{\bath}}
\author{Hendrik~van~Eerten\altaffilmark{\bath}}
\author{Patricia~Schady\altaffilmark{\bath}}
\author{C.~G.~Mundell\altaffilmark{\bath}}
\author{Kate~D.~Alexander\altaffilmark{\northwestern\cierafellow}}
\author{Rodolfo~Barniol Duran\altaffilmark{\cssac}}
\author{Edo~Berger\altaffilmark{\cfa}}
\author{J.~Bolmer\altaffilmark{\mpe}}
\author{Ryan~Chornock\altaffilmark{\ou}}
\author{Deanne~L.~Coppejans\altaffilmark{\northwestern}}
\author{Wen-fai~Fong\altaffilmark{\northwestern}}
\author{Andreja~Gomboc\altaffilmark{\ngslov}}
\author{N\'uria~Jordana-Mitjans\altaffilmark{\bath}}
\author{Shiho~Kobayashi\altaffilmark{\lvjm}}
\author{Raffaella~Margutti\altaffilmark{\northwestern}}
\author{Karl~M.~Menten\altaffilmark{\mpir}}
\author{Re'em~Sari\altaffilmark{\huji}}
\author{Ryo~Yamazaki\altaffilmark{\aoyama}}
\author{V.~M.~Lipunov\altaffilmark{\mossuphys,\mossuastro}}
\author{E.~Gorbovskoy\altaffilmark{\mossuastro}}
\author{V.~G.~Kornilov\altaffilmark{\mossuphys,\mossuastro}}
\author{N.~Tyurina\altaffilmark{\mossuastro}}
\author{D.~Zimnukhov\altaffilmark{\mossuastro}}
\author{R.~Podesta\altaffilmark{\oafa}}
\author{H.~Levato\altaffilmark{\icate}}
\author{D.~A.~H.~Buckley\altaffilmark{\saao}}
\author{A.~Tlatov\altaffilmark{\pulkovo}}
\author{R.~Rebolo\altaffilmark{\iac}}
\author{M.~Serra-Ricart\altaffilmark{\iac}}

\affil{\altaffilmark{\bath}Department of Physics, University of Bath, Claverton Down, Bath, BA2 
7AY, United Kingdom}
\affil{\altaffilmark{\northwestern} Center for Interdisciplinary Exploration and Research in 
Astrophysics (CIERA) and Department of Physics and Astronomy, Northwestern University, Evanston, 
IL 60208, USA}
\affil{\altaffilmark{\cssac}Department of Physics and Astronomy, California State University, 
Sacramento, 6000 J Street, Sacramento, CA 95819, USA}
\affil{\altaffilmark{\cfa}Department of Astronomy, Harvard University, 60 Garden Street, 
Cambridge, MA 02138, USA}
\affil{\altaffilmark{\mpe}Max-Planck-Institut f\"ur extraterrestrische Physik, 
Giessenbachstra{\ss}e, D-85748 Garching, Germany}
\affil{\altaffilmark{\ou}Astrophysical Institute, Department of Physics and Astronomy, 251B 
Clippinger Lab, Ohio University, Athens, OH 45701, USA}
\affil{\altaffilmark{\ngslov}Center for Astrophysics and Cosmology, University of Nova Gorica, 
Vipavska 13, 5000 Nova Gorica, Slovenia}
\affil{\altaffilmark{\lvjm}Astrophysics Research Institute, Liverpool John Moores University, 
IC2, Liverpool Science Park, 146 Brownlow Hill, Liverpool L3 5RF, United Kingdom}
\affil{\altaffilmark{\mpir}Max-Planck-Institut f{\"u}r Radioastronomie, Auf dem H{\"u}gel 69, 
D-53121 Bonn, Germany}
\affil{\altaffilmark{\huji}Racah Institute of Physics, The Hebrew University, Jerusalem 91904, 
Israel}
\affil{\altaffilmark{\aoyama}Department of Physics and Mathematics, Aoyama-Gakuin University, 
Kanagawa 252-5258, Japan}
\affil{\altaffilmark{\mossuphys}M.~V.~Lomonosov Moscow State University, Physics Department, 
Leninskie gory, GSP-1, Moscow, 119991, Russia}
\affil{\altaffilmark{\mossuastro}M.~V.~Lomonosov Moscow State University, Sternberg Astronomical 
Institute, Universitetsky pr., 13, Moscow, 119234, Russia}
\affil{\altaffilmark{\oafa}Observatorio Astronomico Felix Aguilar (OAFA), National University of 
San Juan, San Juan, Argentina}
\affil{\altaffilmark{\icate}Instituto de Ciencias Astronomicas, de la Tierra y del Espacio (ICATE), 
San Juan, Argentina}
\affil{\altaffilmark{\saao}South African Astronomical Observatory, P.O. Box 9, 7935 Observatory, 
Cape Town, South Africa}
\affil{\altaffilmark{\pulkovo}Kislovodsk Solar Station of the Main (Pulkovo) Observatory, RAS, P.O. 
Box 45, ul. Gagarina 100, Kislovodsk, 357700, Russia}
\affil{\altaffilmark{\iac}Instituto de Astrof\'isica de Canarias (IAC), Calle V\'ia L\'actea s/n, 
E-38200 
La Laguna, Tenerife, Spain}

\altaffiltext{\cierafellow}{NHFP Einstein Fellow}

\begin{abstract}
We present comprehensive multiwavelength radio to X-ray observations of GRB~181201A spanning from 
$\approx150$~s to $\approx163$~days after the burst, comprising the first joint ALMA--VLA--GMRT 
observations of a gamma-ray burst (GRB) afterglow. The radio and millimeter-band data reveal a distinct 
signature at $\approx3.9$~days, which we interpret as reverse-shock (RS) emission. Our observations 
present the first time that a single radio-frequency spectral energy distribution can be 
decomposed directly into RS and forward shock (FS) components. We perform detailed modeling of the 
full multiwavelength data set, using Markov Chain Monte Carlo sampling to construct the joint 
posterior density function of the underlying physical parameters describing the RS and FS 
synchrotron emission. We uncover and account for all {discovered} degeneracies in the model 
parameters. 
The joint RS--FS modeling reveals a weakly magnetized ($\sigma\approx3\times10^{-3}$), mildly 
relativistic RS, from which we derive an initial bulk Lorentz factor of $\Gamma_0\approx103$ for 
the GRB jet. Our results support the hypothesis that low-density environments are conducive to the 
observability of RS emission. We compare our observations to other events with strong RS detections 
and find a likely observational bias selecting for longer lasting, non-relativistic RSs. 
We present and begin to address new challenges in modeling posed by the present generation of 
comprehensive, multifrequency data sets.
\end{abstract}

\keywords{gamma-ray burst: general -- gamma-ray burst: individual (GRB\,181201A)}

\section{Introduction}
\label{text:intro}
Long-duration gamma-ray bursts, produced in the core collapse of massive stars, host the most 
energetic and most highly relativistic jets in the universe \citep{mr93,rm94,fkn+97,wb06}. 
However, the nature of the central engine, as well as the mechanism by which these jets are launched 
and collimated, remain poorly understood \citep{mes06}. Probing the nature of the central engine, 
such as magnetars or an accreting black hole, requires foremost determining the energetics and 
composition of the GRB jets themselves \citep{mgt+11,lcj17}. Unfortunately, the forward shock (FS) 
synchrotron `afterglow' emission, produced by the interaction of the jet with the environment, is 
insensitive to the properties of GRB jets; instead, the baryon content, initial bulk Lorentz factor, 
and magnetization of the ejecta can be studied by capturing emission from the reverse shock (RS), 
produced in the jet during its rapid deceleration by the ambient medium 
\citep{spn98,pk00,gk03,gt05,zk05}. 

Strong RSs are expected in ejecta where the magnetic field is dynamically unimportant, and the ratio 
of the magnetic energy density to the bulk kinetic energy $\sigma\lesssim1$. On the other hand, 
highly magnetized outflows are expected to produce weak RS emission \citep{zk05,mkp06,lcj17}. 
The expected signature of RS radiation comprises a characteristic synchrotron spectrum superposed 
on the afterglow emission \citep{kob00}, resulting in a flash beginning in the optical 
and typically lasting a few tens of seconds to minutes, and which rapidly cascades to radio 
frequencies, where it typically lasts a few days \citep{abb+99,sp99a,np04}. 

Recent multiwavelength observations have unequivocally isolated excess radio emission attributable 
to the RS, and a careful decomposition of the observed multifrequency (radio to X-ray) spectral 
energy distribution (SED) at different epochs into FS and RS contributions has allowed the first 
inferences of the Lorentz factor and magnetization of GRB jets 
\citep{lbz+13,pcc+14,vdhpdb+14,lab+16,alb+17,lbm+18,lab+18}. These works have usually provided point 
estimates on the RS parameters with no uncertainty estimates, owing to limitations of the modeling 
process. 

Here, we present multiwavelength observations of GRB~181201A spanning from $\approx150$~s to 
$\approx163$~days after the burst at 35 frequencies from $6\times10^{8}$~Hz to 
$3\times10^{17}$~Hz, including the first joint Giant Meterwave Radio Telescope (GMRT), Karl 
G.~Jansky Very Large Array (VLA), and Atacama Large Millimeter/submillimeter Array (ALMA) data 
set (Section~\ref{text:obs_basic}). Our radio and millimeter-band observations 
reveal an RS spectral peak superposed on the FS emission, which we are able to decompose into RS and 
FS components in a single radio SED for the first time (Section~\ref{text:basic}), while using the 
optical and X-ray observations to constrain the nature of the circumburst environment. 
We fit the entire multiwavelength dataset simultaneously to derive the properties of the RS and FS 
emission, and demonstrate that the two components are consistent with emission from a double-shock 
system (Section~\ref{text:multimodel}). We use this decomposition to derive the initial Lorentz 
factor and magnetization of the ejecta using full statistical modeling. We conclude with a detailed 
discussion of the results, placed in context of other GRBs with multiwavelength RS detections 
(Section~\ref{text:discussion}). Unless otherwise stated, all uncertainties are reported at the 
$1\sigma$ level, and all magnitudes are in the AB system and not corrected for Galactic extinction. We 
employ the $\Lambda$CDM cosmological parameters of $\Omega_m=0.31$, $\Omega_{\Lambda}=0.69$, and 
$H_0=68$\,km\,s$^{-1}$\,Mpc$^{-1}$ throughout.

\section{Observations and Data Analysis}
\label{text:obs_basic}
GRB~181201A was discovered by the IBIS and ISGRI instruments on board the \textit{INTEGRAL} 
satellite on 2018 December 1 at 02:38 UT \citep{gcn23469}. The burst, which lasted $\approx180$~s, 
saturated the telemetry, and hence only a lower limit to its $\gamma$-ray flux is available from 
\textit{INTEGRAL}. The GRB was also detected by the \textit{Fermi} Large Area Telescope (LAT) 
beginning at 02:38:00 UT \citep{gcn23480}, by \textit{Konus}-Wind at 02:39:52.36 UT with a burst 
duration of $\approx172$~s \citep{gcn23495}, and by \textit{Astrosat} with a duration of 
$T_{90}\approx19.2$~s \citep{gcn23501}. We take the \textit{Fermi} and \textit{INTEGRAL} time as 
$T_0$ for this burst, and reference all times hereafter to this $T_0$. The time-averaged 
\textit{Konus}-Wind spectrum in the 20 keV--10 MeV energy range can be fit with a Band model, with 
low-energy photon index $\alpha_{\gamma} = -1.25\pm0.05$, high-energy photon index $\beta = 
-2.73_{-0.12}^{+0.11}$, peak energy $E_{\gamma,\rm peak} = (152\pm6)$~keV, and a fluence of 
$(1.99\pm0.06)\times10^{-4}$~erg\,cm$^{-2}$ \citep{gcn23495}. 

The optical afterglow was discovered in MASTER \citep{lkg+10} Global Robotic Net observations 
\citep{gcn23470} 10\,s after the \textit{INTEGRAL} alert and later by the \Swift\ UV/optical 
telescope \citep[UVOT;][]{rkm+05,gcn23499}. Spectroscopic 
observations with the Very Large Telescope at the European Southern Observatory yielded a 
redshift of $z=0.450$ based on identification of the Mg~II doublet \citep{gcn23488}. 
At this redshift, the \textit{Konus}-Wind spectrum implies an isotropic-equivalent $\gamma$-ray 
energy of $E_{\gamma,\rm iso}=(1.20\pm0.04)\times10^{53}$~erg, 
This value is consistent with the GRB $E_{\rm peak}$-$E_{\gamma,\rm iso}$ relation 
\citep{ama06}, suggesting that the prompt $\gamma$-ray emission in GRB~181201A is typical among the 
GRB population.

\subsection{X-ray: Swift/XRT}
The \Swift\ X-ray Telescope \citep[XRT,][]{bhn+05} began observing GRB\,181201A 0.16~days after the 
burst, and detected a fading X-ray afterglow at RA = $21^{\rm h}17^{\rm m}11{\rlap.}^{\rm 
s}$20, Dec = -12$^\circ$\,37\arcmin\,50.9\arcsec\ (J2000), with an uncertainty radius of 
1.4\arcsec\ (90\% 
containment)\footnote{\url{http://www.swift.ac.uk/xrt_positions/00020848/}}. XRT continued 
observing the afterglow for $20$\,d in photon counting mode. The GRB entered a Sun constraint 
for \Swift\ at $\approx21$~days after the burst and was dropped from the \Swift\ schedule. We 
obtained 4.4~ks of \Swift\ ToO observations on 2019 April 7 ($\approx 127$~days), when it was 
visible again. The X-ray afterglow was weakly detected. The data were automatically calibrated 
using the XRT pipeline and incorporated into the on-line light curve. We use a photon index of 
$\Gamma_{\rm X}\approx1.77$ and an unabsorbed counts-to-flux conversion rate from the \Swift\ 
website of $4.75\times10^{-11}$\,erg\,\pcmsq\,ct$^{-1}$ to convert the 0.3--10\,keV count rate 
light 
curve\footnote{Obtained from the \Swift\ website at 
\url{http://www.swift.ac.uk/xrt_curves/00020848} 
and re-binned to a minimum signal-to-noise ratio per bin of 8.} to flux density at 1\,keV for 
subsequent analysis.

\begin{deluxetable}{clccc}
\tabletypesize{\scriptsize}
\tablecaption{\Swift\ UVOT Observations of GRB\,181201A \label{tab:data:UVOT}}
\tablehead{
\colhead{Mid-time,} & \colhead{UVOT} & \colhead{Flux density} & \colhead{Uncertainty}
& \colhead{Detection?}\\ 
\colhead{$\Delta t$ (d)} & \colhead{band} & \colhead{(mJy)} & \colhead{(mJy)} & \colhead{(1=Yes)}}

\startdata
0.176 & \textit{uw2} & $6.19\times10^{-1}$ & $2.92\times10^{-2}$ & 1 \\
0.246 & \textit{uw2} & $4.61\times10^{-1}$ & $2.17\times10^{-2}$ & 1 \\
0.302 & \textit{uw2} & $3.31\times10^{-1}$ & $1.88\times10^{-2}$ & 1 \\
\ldots & \ldots & \ldots
\enddata
\tablecomments{This is a sample of the full table available on-line.}
\label{tab:data:uvot}
\end{deluxetable}

\begin{deluxetable*}{ccccccccc}
\tabletypesize{\scriptsize}
\tablecaption{Optical and Near-IR Observations of GRB\,181201A \label{tab:data:opt}}
\tablehead{
\colhead{$\Delta t$} & \colhead{Observatory} & \colhead{Instrument} &
\colhead{Filter} & \colhead{Frequency} & \colhead{Flux density} & 
\colhead{Uncertainty} & \colhead{Detection?} & \colhead{Reference}\\ 
\colhead{(d)} & & & & \colhead{(Hz)} & \colhead{(mJy)} & \colhead{(mJy)} & \colhead{1=Yes} }

\startdata
$1.81\times10^{-3}$ & MASTER-OAFA & MASTER & \textit{CR} & $4.67\times10^{14}$ & $1.25\times10^{1}$ 
& $1.27\times10^{-1}$ & 1 & This work \\
$2.42\times10^{-3}$ & MASTER-OAFA & MASTER & \textit{CR} & $4.67\times10^{14}$ & $2.85\times10^{1}$ 
& $1.58\times10^{-1}$ & 1 & This work \\
$3.46\times10^{-1}$ & iTelescope & T31 & \textit{R} & $4.67\times10^{14}$ & $6.52\times10^{-1}$ & 
$1.83\times10^{-2}$ & 1 & \cite{gcn23475} \\
\ldots & \ldots & \ldots & \ldots & \ldots & \ldots & \ldots & \ldots & \ldots
\enddata
\tablecomments{The data have not been corrected for Galactic extinction. This is a sample of the 
full table available on-line.}
\end{deluxetable*}

\subsection{UV, optical, and near-IR}
We derived aperture photometry from pipeline processed UVOT images downloaded from the \Swift\ 
data center\footnote{\url{http://www.swift.ac.uk/swift\_portal}} with the {\sc uvotproduct} (v2.4) 
software. We used a circular source extraction region with a 3.5\arcsec\ radius, followed by an 
aperture correction based on the standard 5\arcsec\ aperture for UVOT photometry \citep{pbp+08}. We 
present our UVOT measurements in Table~\ref{tab:data:UVOT}. 

Four MASTER-net observatories (MASTER-Kislovodsk, MASTER-IAC, MASTER-SAAO, and MASTER-OAFA) 
participated in rapid-response and follow-up observations of GRB181201A, beginning with 10~s 
exposures at 02:40:30 UT on 2019 December 1,  $\approx150$~s after the burst and 10~s after the 
\textit{INTEGRAL} alert in $BVRI$ and Clear bands\footnote{MASTER Clear band ($CR$) magnitudes are 
best described by the ratio, $CR = 0.8R + 0.2B$.} \citep{gcn23470}. The MASTER auto-detection 
system \citep{lkg+10,lvg+19} discovered a bright optical afterglow at RA = $21^{\rm h}17^{\rm 
m}11{\rlap.}^{\rm s}$20, Dec = 
-12$^\circ$\,37\arcmin\,51.4\arcsec\ (J2000) with unfiltered magnitude $\approx13.2$~mag (Vega). 
MASTER follow-up observations continued until $\approx3$~days after the burst. We carried out 
astrometry and aperture photometry (using a $\approx2.5\arcsec$ aperture radius) for observations 
taken at each telescope separately, calibrated to 6 nearby SDSS reference stars. We verified our 
photometry by measuring the brightness of comparison stars with similar brightness as the afterglow, 
and incorporated the derived systematic calibration uncertainty into the photometric uncertainty 
\citep{tlm+17,lvg+19}.

We observed the afterglow of GRB~181201A with the Gamma-Ray Burst Optical/Near-Infrared Detector 
(GROND) on the MPI/ESO 2.2m telescope in La Silla in Chile beginning 0.91~days after the burst in 
$g^{\prime}r^{\prime}i^{\prime}z^{\prime}JHK$ filters.  We calibrated the data with tools and 
methods standard for GROND observations and detailed in \citet{kkysg+08}. 
We performed aperture photometry calibrated against PanSTARRS field stars in the 
$g^{\prime}r^{\prime}i^{\prime}z^{\prime}$ filters and against the 2MASS catalog in $JHK$. 

We observed the afterglow with the Ohio State Multi-Object Spectrograph \citep[OSMOS;][]{msd+11} 
on the 2.4 m Hiltner telescope at MDM Observatory on 2018 December 04, with 120~s exposures each in 
the \textit{BVRI} bands. We obtained two later epochs of imaging using the Templeton detector on the 
1.3m McGraw-Hill telescope on the nights of 2018 December 9 and 10, obtaining $4\times300$~s each in 
$r^{\prime}$ and $i^{\prime}$ bands. Observations on December 10 were taken at a high airmass of 
1.91 and under poor seeing. We performed aperture photometry with a 2\arcsec\ aperture on the 
Hiltner and McGraw-Hill images, referenced to SDSS. 

We obtained Liverpool Telescope \citep[LT;][]{ssr+04} imaging with the infrared/optical camera 
(IO:O) at $\approx2.7$~days, comprising $3\times120$~s exposures in 
$g^{\prime}r^{\prime}i^{\prime}$ 
bands. We performed aperture photometry using the Astropy Photutils package \citep{bsr+16} 
calibrated to SDSS. We also collected all optical photometry for this event published in GCN 
circulars. We present our combined optical and NIR data set in Table~\ref{tab:data:opt}.

\begin{deluxetable}{ccccc}
 \tabletypesize{\footnotesize}
 \tablecolumns{5}
 \tablecaption{Radio and millimeter observations of GRB~181201A}
 \tablehead{   
   \colhead{Telescope} &
   \colhead{Frequency} &
   \colhead{Time} &
   \colhead{Flux density} &
   \colhead{Uncertainty}\\
   \colhead{} &
   \colhead{(GHz)} &
   \colhead{(days)} &
   \colhead{(mJy)} &
   \colhead{($\mu$Jy)}
   }
 \startdata 
ALMA &$97.5$ &$0.885$ &$3.41$ &$23$ \\
ALMA &$96.4$ &$1.92$ &$1.99$ &$25$ \\
VLA &$2.73$ &$2.93$ &$0.478$ &$70$ \\
\ldots & \ldots & \ldots & \ldots & \ldots
\enddata
\tablecomments{This is a sample of the full table available on-line.}
\label{tab:data:radio}
\end{deluxetable}

\subsection{Millimeter: ALMA}
We obtained 5 epochs of ALMA Band 3 (3\,mm) observations of GRB~181201A spanning 0.88~days to 
29.8~days after the burst through program 2018.1.01454.T (PI: Laskar). The observations utilized 
two 4 GHz-wide base-bands centered at 91.5 and 103.5~GHz, respectively, in the first epoch, and at 
90.4 and 102.4~GHz in subsequent epochs. One of J2258-2758, J2148+0657, and J2000-1748 was used for 
bandpass and flux density calibration, while J2131-1207 was used for complex gain calibration in 
all epochs. We analyzed the data using the Common Astronomy Software Application 
\citep[][CASA]{mws+07} and the afterglow was well-detected in all epochs. 
We performed one round of phase-only self-calibration in the first two epochs, where the 
deconvolution residuals indicated the presence of phase errors. We list the results of our ALMA 
observations in Table \ref{tab:data:radio}.

\subsection{Centimeter: VLA}
We observed the afterglow using the VLA starting 2.9\,d after 
the burst through programs 18A-088 and 18B-407 (PI: Laskar). We detected and tracked the flux 
density of the afterglow from 1.2\,GHz to 37\,GHz over multiple epochs until $\approx164$\,d after 
the burst. We used 3C48 as the flux and bandpass calibrator and J2131-1207 as complex gain 
calibrator. Where phase errors appeared to be present after deconvolution, we performed phase-only 
self-calibration in epochs with sufficient signal-to-noise. The effect of self-calibration impacted 
the measured flux density by $\lesssim15\%$ and only at the highest frequencies ($\gtrsim15$~GHz). 
We carried out data reduction using CASA, and list the results of our VLA observations in Table 
\ref{tab:data:radio}.

\subsection{Meter: uGMRT}
We observed the afterglow using the upgraded Giant Meterwave Radio Telescope (uGMRT) through 
program 
35\_065 (PI: Laskar) starting 12.5\,d and 13.5\,d after the burst in Bands 4 (center 
frequency 550~MHz) and 5 (center frequency 1450~MHz), respectively. The observations utilized the 
full 400~MHz bandwidth of the uGMRT Wide-band Backend (GWB) system. We used 3C48 for 
bandpass and flux density calibration, and J2131-1207 for complex gain calibration. We carried out 
data reduction using CASA, and list the results of our uGMRT observations in Table 
\ref{tab:data:radio}.

\section{Basic Considerations}
\label{text:basic}
We now interpret the X-ray to radio afterglow observations in the context of synchrotron emission 
arising from electrons accelerated to a non-thermal power-law distribution, $N_{\rm 
\gamma_{\rm e}}\propto \gamma_{\rm e}^{-p}$, in relativistic shocks produced by the interaction of 
the GRB jet with the circum-burst environment \citep{spn98}. We assume the energy density of 
accelerated particles and that of the magnetic field downstream of the shock to be a fraction 
$\epse$ and $\epsb$, respectively, of the shock kinetic energy density for a given total 
ejecta kinetic energy, $E_{\rm K,iso}$.
The observed spectral energy distribution (SED) of each 
synchrotron component is described by three break frequencies (the self-absorption 
frequency, \nua, the characteristic frequency, \numax, and the cooling frequency, \nuc), and the 
flux density normalization, \fnumax\ \citep{gs02}. The evolution of these 
quantities with time depends upon the radial density profile of the circum-burst medium, $\rho = 
AR^{-k}$, for which we consider two standard possibilities: a constant density ($k=0$, $A\equiv\dens 
m_{\rm p}$; ISM) model \citep{spn98} 
and a wind-like ($k=2$, $A\equiv5\times10^{11}$g\,cm$^{-1}\Astar$) environment \citep{cl00}. 
We compute the resulting synchrotron spectra using the weighting prescription described in 
\cite{lbt+14} to generate smooth light curves at all observed frequencies.

\subsection{Radio/mm: presence of additional component}
\label{text:4dsed}
\begin{figure}
  \centering
  \includegraphics[width=\columnwidth]{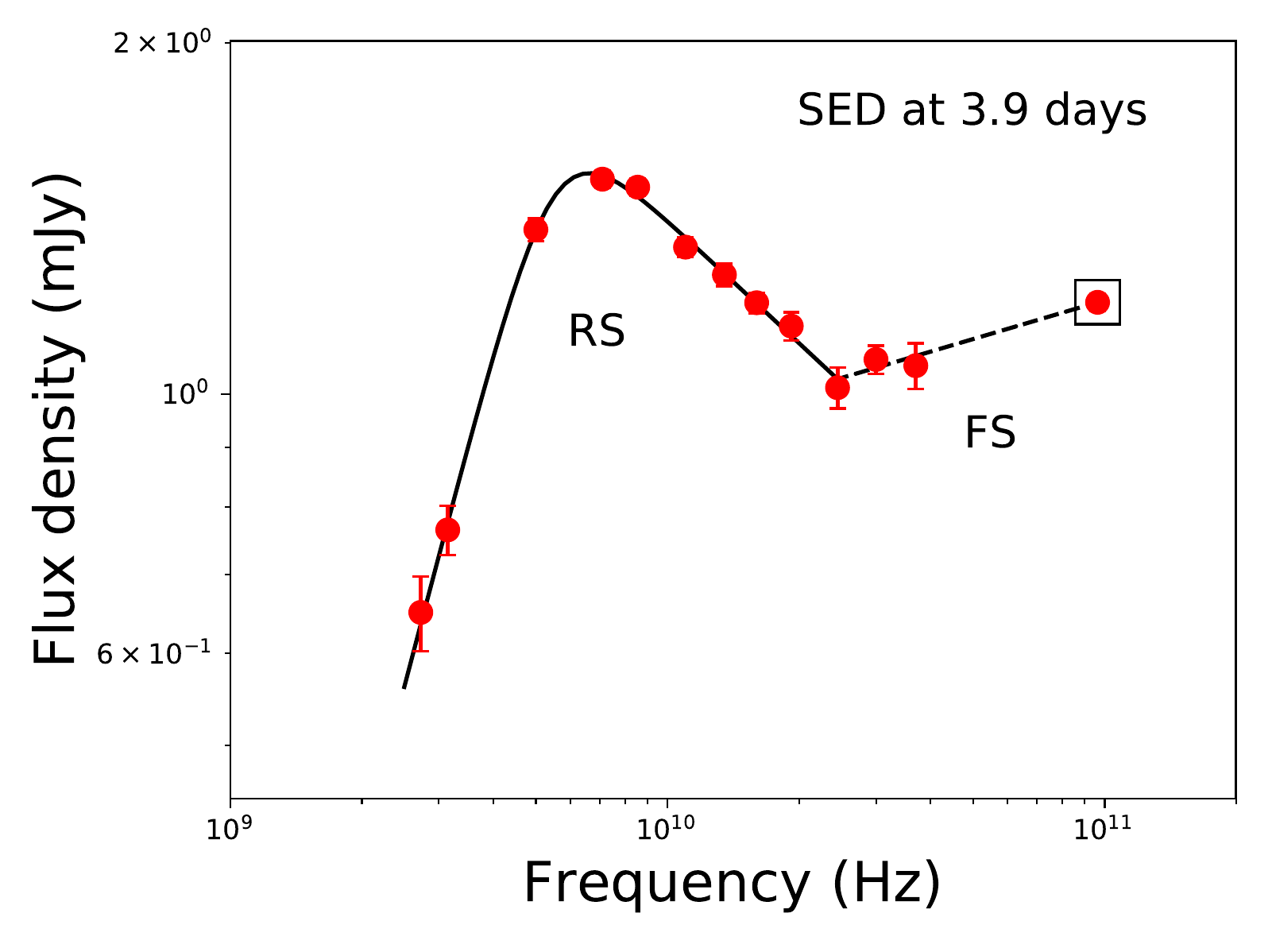}
  \caption{Radio and mm-band SED of GRB~181201A at 3.9~days post-burst from our VLA and ALMA 
programs, with a broken power law fit at 3--24 GHz (solid) and a single power law at high 
frequencies (dashed). The cm-band data spanning 3--24 GHz exhibit a self-absorbed spectral peak, 
which we attribute to emission from a reverse shock. The VLA observations above 24\,GHz and the
the ALMA data (highlighted by a black box) constrain the high-frequency rising spectrum with 
index $\beta=0.09\pm0.02$, and form a distinct emission component, which we associate with 
underlying forward shock emission (Section~\ref{text:4dsed}).}
\label{fig:4dsed}
\end{figure}

One of the most striking features of GRB~181201A revealed by our VLA and ALMA observations is a 
spectral bump in the cm to mm-band SED at 3.9~days (Fig.~\ref{fig:4dsed}). The feature spans 3--24 
GHz, and can be fit with a smoothly broken power law, with low-frequency 
index\footnote{We use the convention $F_{\nu}\propto t^{\alpha}\nu^{\beta}$ throughout.}, 
$\beta_1 = 1.43\pm0.23$, high-frequency index, $\beta_2 = -0.35\pm0.03$, break frequency, $\nu_{\rm 
peak}=5.5\pm0.4$~GHz, and peak flux density, $F_{\nu,\rm peak}=1.48\pm0.06$\,mJy. 
The steeply rising spectrum below the spectral peak suggests 
the underlying emission is self-absorbed\footnote{The deviation of this low-frequency power law 
from the expected value of $\beta\approx2$ arises from a contribution from an underlying 
$\nu^{1/3}$ power law extending down from the mm band (Section~\ref{text:multimodel}).}. Fitting the 
VLA cm-band data above 24~GHz and the ALMA observations at the same epoch with a single power law 
model yields $\beta_{\rm 24-ALMA}=0.09\pm0.02$. 

The radio and millimeter SED at 3.9~days clearly exhibits two distinct components. The simplest 
explanation for multiple components in the radio is to attribute one to a RS and the other to a FS. 
Since the RS emission is expected to peak at lower frequencies than the FS and the RS spectrum is 
expected to cut off steeply above the RS cooling frequency \citep{kob00}, the lower-peaking 
component is more naturally associated with RS emission. 

\begin{figure}
  \centering
  \includegraphics[width=\columnwidth]{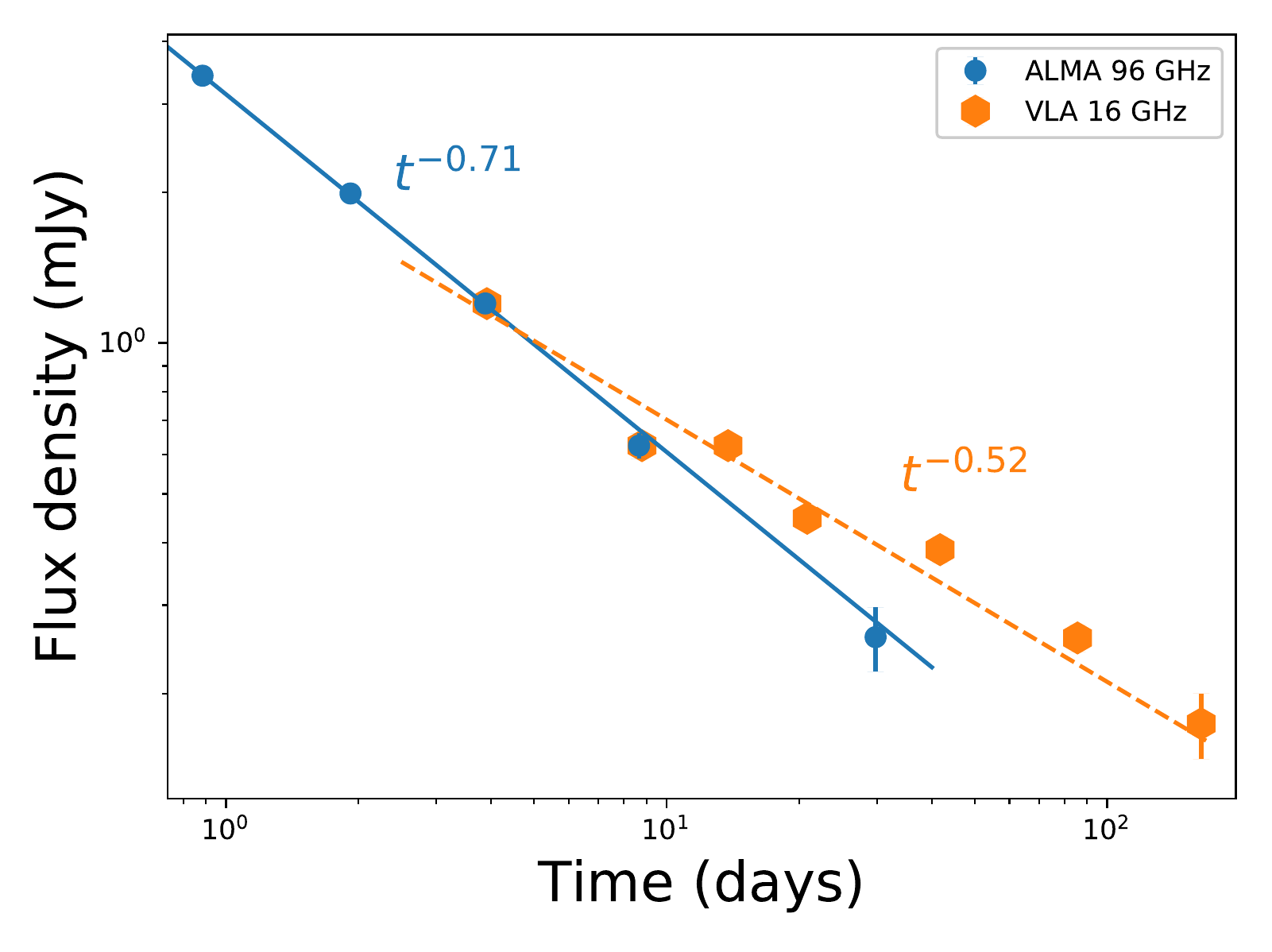}
  \caption{ALMA Band 3 (3\,mm) light curve of GRB~181201A from $\approx$ 0.9~days to 30~days 
after the burst (blue circles) with a power law model fit (blue line; $\alpha = -0.71\pm0.01$), 
along with VLA 16~GHz light curve (orange) and corresponding power law fit for comparison. 
The steep light curve decay indicates a wind-like pre-explosion environment 
(Section~\ref{text:densityprofile}).}
\label{fig:ALMAlc_pl}
\end{figure}

The ALMA mm-band light curve can be fit with a single power law, with decay rate 
$\alpha_{\rm ALMA}=-0.71\pm0.01$ (Fig.~\ref{fig:ALMAlc_pl}). From the radio SED at 3.9~days, we 
know that $\numf > \nu_{\rm ALMA}$ at this time. In this spectral regime, the expected rise rate 
for FS emission is $t^{1/2}$ in the ISM environment and $t^0$ in the wind environment, both of 
which are incompatible with the observed mm-band light curve. Thus, the mm-band light curve at 
$<4$~days originates from a separate component. 
The shallowest decline rate for an RS light curve at $\nu\gtrsim\numr$ is $\alpha\approx-2.1$ 
for a relativistic RS and $\alpha-1.5$ for a non-relativistic RS, in each case corresponding to 
a wind environment with $p\approx2$. Both of these are steeper than the observed mm-band light 
curve. However, the sum of the contributions from the two emission components (RS and FS) may yield 
the observed shallower decay. We investigate this further in Section \ref{text:mmexcess}.

For the remainder of this article, we associate the low- and high-frequency components with 
emission from the RS and the FS, respectively, and consider other scenarios in Section 
\ref{text:other_scenarios}. 
In this framework, our observations of GRB~181201A provide the first instance where a single radio 
SED can be instantaneously and cleanly decomposed into RS and FS components, with the RS dominating 
at the lower frequency bands, and the underlying FS emission clearly visible as an optically thin, 
rising spectrum at higher frequencies. We note that a more detailed characterization of these 
components, including the determination of the precise locations of their spectral break 
frequencies, requires a multi-band model fit, which we undertake in Section~\ref{text:multimodel}.

\subsection{Circum-burst density profile}
\label{text:densityprofile}
\begin{figure}
  \centering
  \includegraphics[width=\columnwidth]{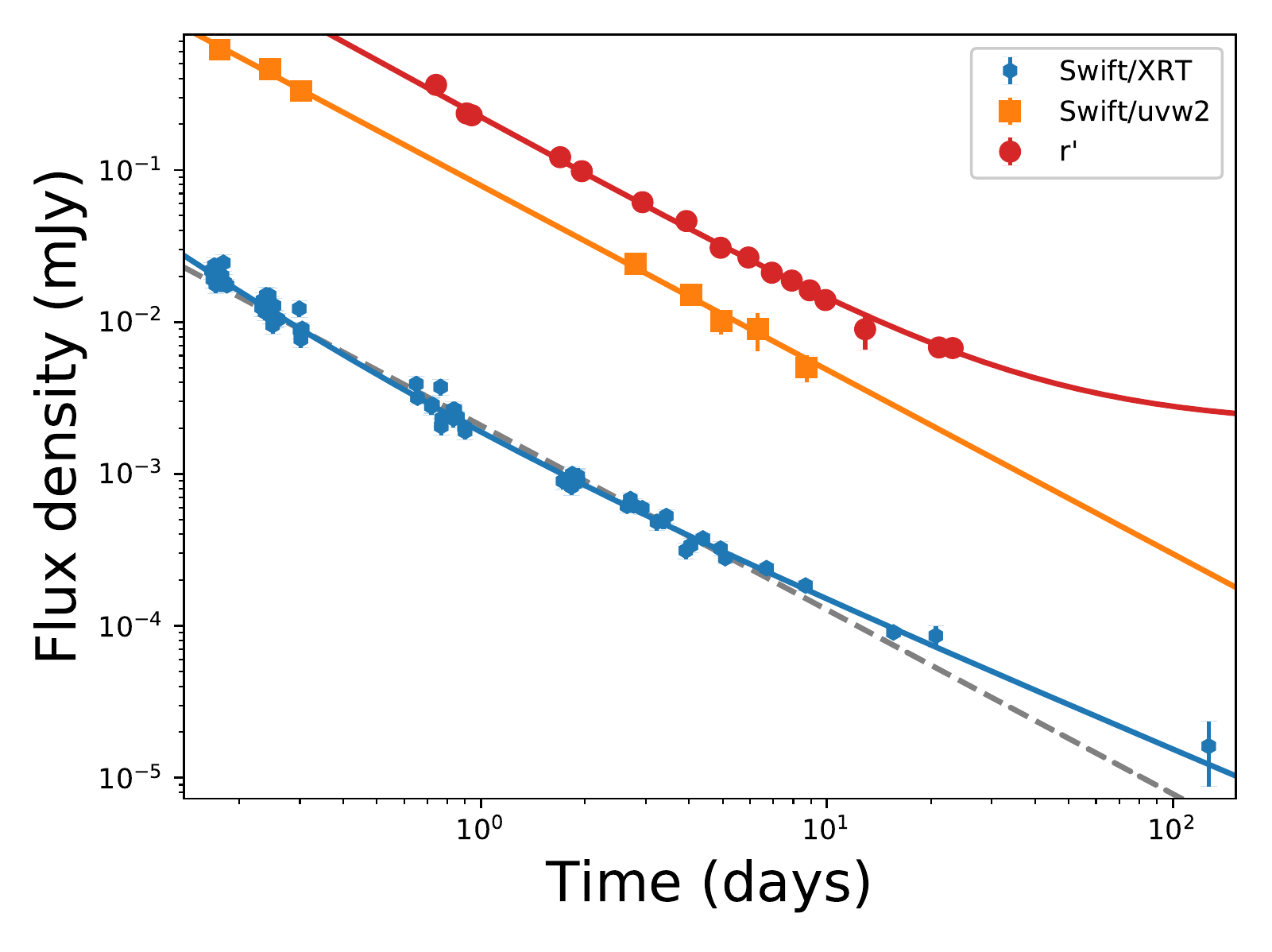}
  \caption{Optical $r^{\prime}$-band, \Swift/UVOT \textit{uvw2}-band, and \Swift/XRT 1~keV light 
curves of GRB~181201A, together with a power-law-plus-constant fit (red; $r^{\prime}$-band), 
power-law fit at \textit{uvw2}-band, and power-law (grey, dashed) and broken power law fit (blue; 
X-ray). The excess emission in $r^{\prime}$-band at $\gtrsim10$~days may indicate contamination from 
the underlying host galaxy. 
}
\label{fig:XRToptlc_pl}
\end{figure}

The identification of the FS component now allows us to constrain the circum-burst density profile. 
We note that there is no evidence for a jet break 
until $\approx21$~days, as the X-ray light curve declines as a single power law from 
$\approx0.15$~days to $\approx21$~days ($\alpha_{\rm X} = -1.22\pm0.01$; 
Fig.~\ref{fig:XRToptlc_pl}). The rising radio spectrum to $\approx90$~GHz at $3.9$~days indicates 
that $\numf\gtrsim90$~GHz at this time. The relatively shallow spectral index of $\beta\approx0.1$ 
compared to the expected index of $\beta=1/3$ may result from the smoothness and potential proximity 
of the $\numf$ break \citep{gs02}.

Because $\numf\gtrsim90$~GHz at 3.9~days, the FS peak flux density, 
$\fnumf\gtrsim F_{\rm 3\,mm}(3.9~{\rm days})\approx1.2$~mJy. In the ISM environment, this peak flux 
density, $F_{\rm p}$, remains fixed and must appear at a lower cm-band frequency, $\nu_{\rm cm}$, 
at a later time given by, 
\begin{equation}
t_{\rm p} = \left(\frac{\nu_{\rm ALMA}}{\nu_{\rm cm}}\right)^{2/3}  \left(\frac{F_{\rm p}}{F_{\rm 
ALMA}}\right)^2.  
\end{equation}
This implies a flux density of $F_{\rm p}\gtrsim1.2$~mJy at 
$t_{\rm p} \gtrsim13.2$~days at 16~GHz, which is a factor of 2 brighter than the observed 16~GHz 
light curve at the corresponding time. Indeed, all cm-band light curves are inconsistent with this 
extrapolation, and thus inconsistent with an ISM-like environment. 

\begin{figure}
  \centering
  \includegraphics[width=\columnwidth]{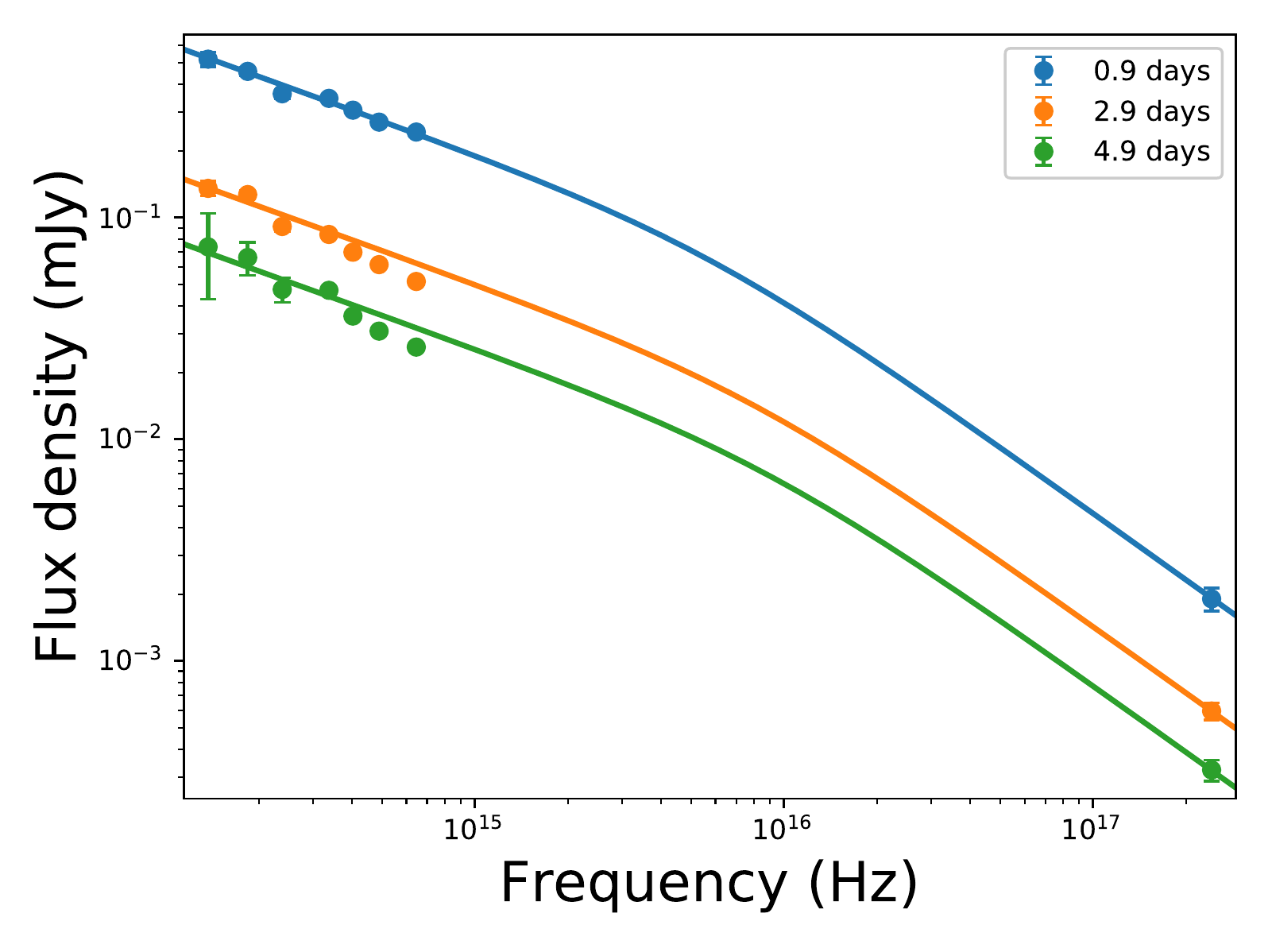}
  \caption{UV to X-ray SEDs at 0.9, 2.9, and 4.9~days, together with broken power law model fits 
(lines). 
  The spectral break suggests $\nuopt<\nucf<\nux$ at this time (Section~\ref{text:densityprofile}. }
\label{fig:optxrtsed}
\end{figure}

We now show that the optical to X-ray data are also inconsistent with an ISM environment, and 
indicate a wind-like circum-burst medium.
The GROND SED at 0.91~days corrected for Galactic extinction of $A_{\rm V} = 0.17$~mag \citep{sf11} 
can be fit with a single power law, with index $\beta_{\rm opt} = -0.49\pm0.04$ 
(Fig.~\ref{fig:optxrtsed}). On the other hand, the spectral index between the GROND $K$-band and the 
X-rays is steeper, $\beta_{\rm NIR,X}=-0.75\pm0.02$, indicating that a spectral break lies between 
the optical/NIR and X-rays. The most natural explanation for these observations is $\numf<\nu_{\rm 
opt}<\nucf<\nux$. 
Fitting the optical and X-ray SED at 0.9~days, 2.9~days, and 4.9~days with broken power law 
models\footnote{$F_{\nu} = F_{\rm b} \left[
 \frac{1}{2} \left(\frac{\nu}{\nu_{\rm b}}\right)^{-y\beta_1} + 
 \frac{1}{2} \left(\frac{\nu}{\nu_{\rm b}}\right)^{-y\beta_2} 
 \right]^{-1/y}.$ We fix $y=3$ in this fit to emulate a moderately smooth break.
} 
fixing the low frequency index to 
$\beta_{\rm opt}=-0.5$ and the high frequency index to $\beta_{\rm X}=-1.0$, we find 
a break frequency, $\nu_{\rm b}=(5.7\pm1.5)\times10^{15}$~Hz, $(8.1\pm2.0)\times10^{15}$~Hz,
and $(9.1\pm3.3)\times10^{15}$~Hz, respectively (Fig.~\ref{fig:optxrtsed}). 
The rising break frequency is roughly consistent
with the evolution of $\nucf\propto t^{1/2}$ in a wind-like environment. 
We caution that the inferred location of the break is only indicative,
as it is degenerate with the spectral slopes and break smoothness. 

For $\numf<\nuopt<\nuc$, we expect a spectral index of $\beta=(1-p)/2$. Combined with the observed
GROND SED at 0.9~days, this implies $p\approx2$, which yields a light curve decay rate of 
$\alpha_{\rm opt}\approx-1.25$ for the wind environment and $\alpha_{\rm opt}\approx-0.75$ for the  
ISM environment. The observed decay rate in the \Swift/UVOT $uvw2$-band and in 
ground-based optical $r^{\prime}$-band observations at $\approx0.2$--8.7~days is $\alpha_{\rm UV} = 
-1.21\pm0.02$, consistent with the wind case and inconsistent with an ISM 
environment (Fig.~\ref{fig:XRToptlc_pl}).

Above $\nucf$, we expect a decay rate of $\alpha=(2-3p)/4\approx-1$. A power law fit to the 
X-ray light curve yields a decay rate of $\alpha_{\rm X} = -1.212 \pm 0.013$; however, 
the light curve after $\approx6$~days is consistently above the single power law fit. Adding a 
temporal break at $t_{\rm b}\approx0.4$~days yields a better fit with $\alpha_{X,1}=-1.7\pm0.1$ and 
$\alpha_{X,2}=-0.96\pm0.02$. 
The steep early decay may imply that the cooling break is not far below the X-ray band, and we 
verify this with our full multi-wavelength model in Section~\ref{text:MCMC}. On the other hand, the 
late-time excess is unusual, and we discuss this further in Section~\ref{text:xrayexcess}.

The measured X-ray photon index of $\Gamma_{\rm X}=1.77\pm0.06$ implies $\beta_{\rm 
X}=-0.77\pm0.06$, 
which is not consistent with $\beta=-p/2\approx-1$ expected for $p\approx2$. We note that the 
Klein-Nishina (KN) correction may yield a spectral index of $\beta=-3(p-1)/4\approx-0.75$ in some 
regimes \citep{nas09}, and it is possible that this may impact the X-ray spectrum but not the 
light curve\footnote{When $p\approx2$, the effect of KN corrections on the evolution of \nuc\ with 
time, and hence on the light curve above $\nuc$, is small. 
However, the spectral corrections remain, and in particular, we may have $\nuc < 
\widehat{\nu}_{\rm c} < \nux$, which would result in a spectral index of $\beta\approx-0.75$ 
\citep{nas09}. Here $\widehat{\nu}_{\rm c}/\nuc = (\widehat{\gamma}_{\rm c}/\gamma_{\rm c})^2$, 
$\widehat{\gamma}_{\rm c} = \gamma_{\rm self}^3/\gamma_{\rm c}^2$, $\gamma_{\rm self} = (B_{\rm 
QED}/B)^{1/3}$, $B$ is the post-shock magnetic field, and $B_{\rm QED}$ is the quantum critical 
field.}. We defer a detailed discussion of the KN corrections to future work. 

To summarize, the radio SED at 3.9~days and the mm-band light curve indicate the presence of a 
separate 
emission component than that responsible for the X-ray and optical emission. We interpret this 
low-frequency component as RS emission. 
Furthermore, we have described several observations that favor a wind-like circum-burst 
environment: 
(i) the high mm-band flux density at $4$~days that does not appear in the lower frequency radio 
light curve, (ii) a slowly rising break between the optical and X-rays, and 
(iii) the optical light curve, which is consistent with $\numf<\nuopt<\nuc$ in a wind environment. 
We therefore focus on the wind environment in the remainder of this article.

\subsection{Afterglow onset}
\label{text:optrise1}
The first two MASTER observations at 156 and 209~s after the burst (156~s and 
209~s, respectively) reveal a rapid brightening, $\alpha\approx2.8$. Such 
brightenings have only been seen in a few other optical afterglows \citep{mvm+07,mkm+10,lyz+10}, 
and 
have been interpreted as onset of the afterglow, where the increase in flux density corresponds to 
an on-going transfer of energy from the ejecta to the forward shock, ultimately resulting in the 
deceleration of the ejecta \citep{rm92,mr93,sp99,kps99,kz07}. Unfortunately, our data do not 
capture the time of the peak in the optical bands. As the FS is not fully set up during this 
onset period, we do not include these points in our multiwavelength modeling 
(Section~\ref{text:multimodel}). We discuss these data further in Section~\ref{text:optrise2}.

\begin{deluxetable}{lrr}
\tabletypesize{\scriptsize}
\tablecaption{FS and RS Model Parameters \label{tab:params}}
\tablehead{
\colhead{Parameter} & \colhead{Best fit} & \colhead{MCMC\tablenotemark{a}}
}
\startdata
\multicolumn{3}{c}{Forward Shock}\\
$p$      & $2.11$ & $2.11\pm0.01$\\[2pt]
$\log\epse$  & $-0.43$ & $-0.39^{+0.12}_{-0.17}$\\[2pt]
$\log\epsb$  & $-2.01$ & $-2.20^{+0.42}_{-0.78}$\\[2pt]
$\log\Astar$ & $-1.72$ & $-1.66^{+0.23}_{-0.13}$\\[2pt]
$\log(\EKiso/$erg) & $53.3$ & $53.41^{+0.13}_{-0.16}$\\[2pt]
$\log(\AV/$mag) & $-1.83$ & $-1.84^{+0.14}_{-0.18}$\\[2pt]
$\log\nuaf$  (Hz)  & $7.74$\tablenotemark{b} & \ldots\\[2pt]
$\log\numf$  (Hz)  & $12.7$ & \ldots\\[2pt]
$\log\nucf$  (Hz)  & $16.4$ & \ldots\\[2pt]
$\log\fnumf$ (mJy) & $0.62$ & \ldots \\[2pt]
\hline\\[-4pt]
\multicolumn{3}{c}{Reverse Shock}  \\[2pt]
$\log(\nuar/$Hz)   & $12.1$ & $12.11\pm0.06$ \\[2pt]
$\log(\numr/$Hz)   & $8.6$  & $8.64^{+1.01}_{-0.45}$\\[2pt]
$\log(\nucr/$Hz)   & $14.7$ & $14.65^{+0.25}_{-0.22}$   \\[2pt]
$\log(\fnumr/$mJy) & $5.2$  & $5.16^{+0.26}_{-0.56}$ \\[2pt]
$g$                & 2.72   & $2.74^{+0.38}_{-0.31}$
\enddata
\tablecomments{Frequencies and flux densities are calculated at 1~day (FS) and $10^{-2}$~days (RS).
}
\tablenotetext{a}{Summary statistics from the marginalized posterior density distributions, with 
median and $\pm34.1\%$ quantiles (corresponding to $\pm1\sigma$ for Gaussian distributions; 
Fig.~\ref{fig:corner}).}
\tablenotetext{b}{This frequency is not directly constrained by the data.}
\end{deluxetable}

\begin{figure*} 
 \begin{tabular}{cc}
  \includegraphics[width=0.47\textwidth]{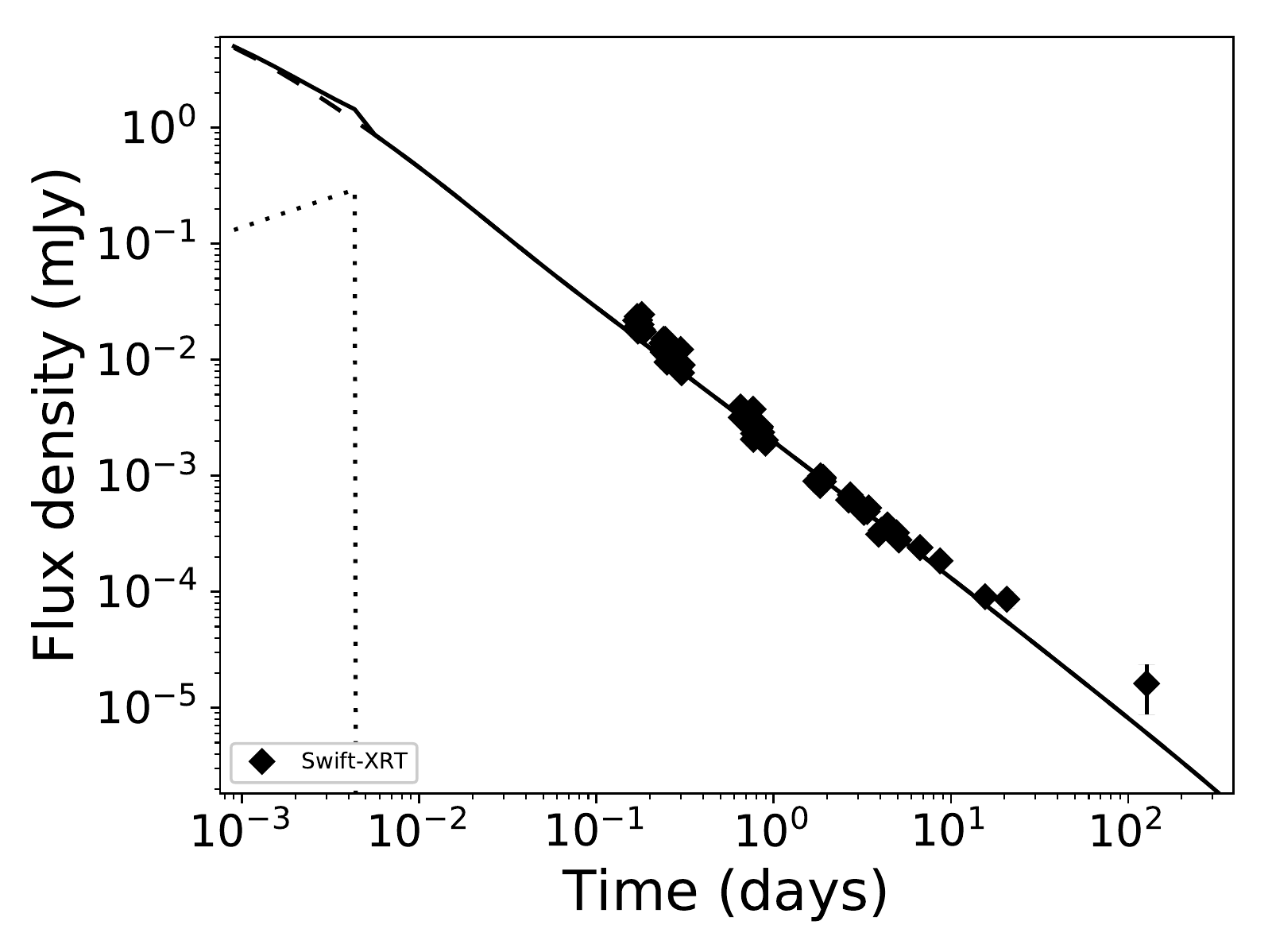} &
  \includegraphics[width=0.47\textwidth]{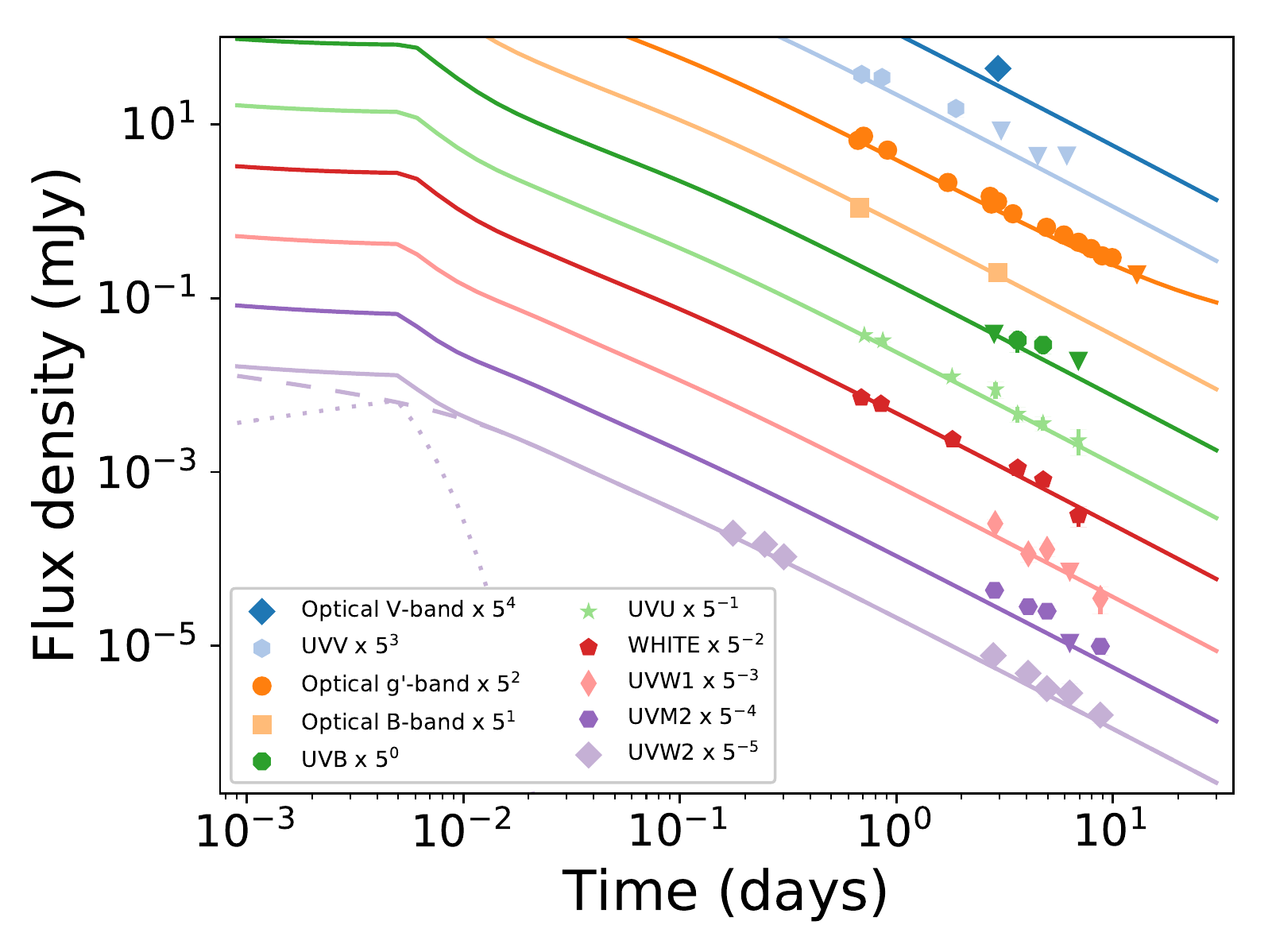} \\ [-3pt]
  \includegraphics[width=0.47\textwidth]{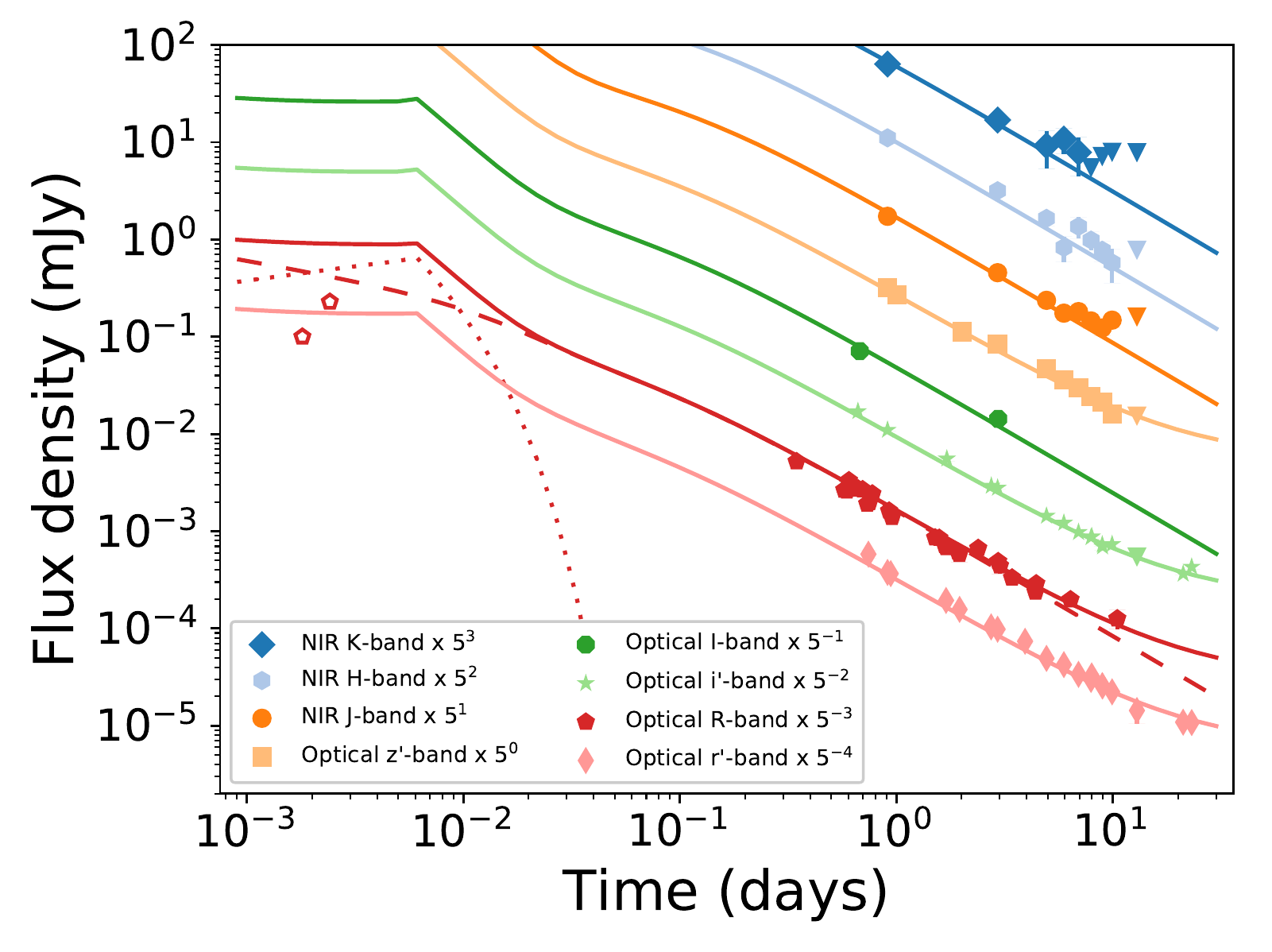} &
  \includegraphics[width=0.47\textwidth]{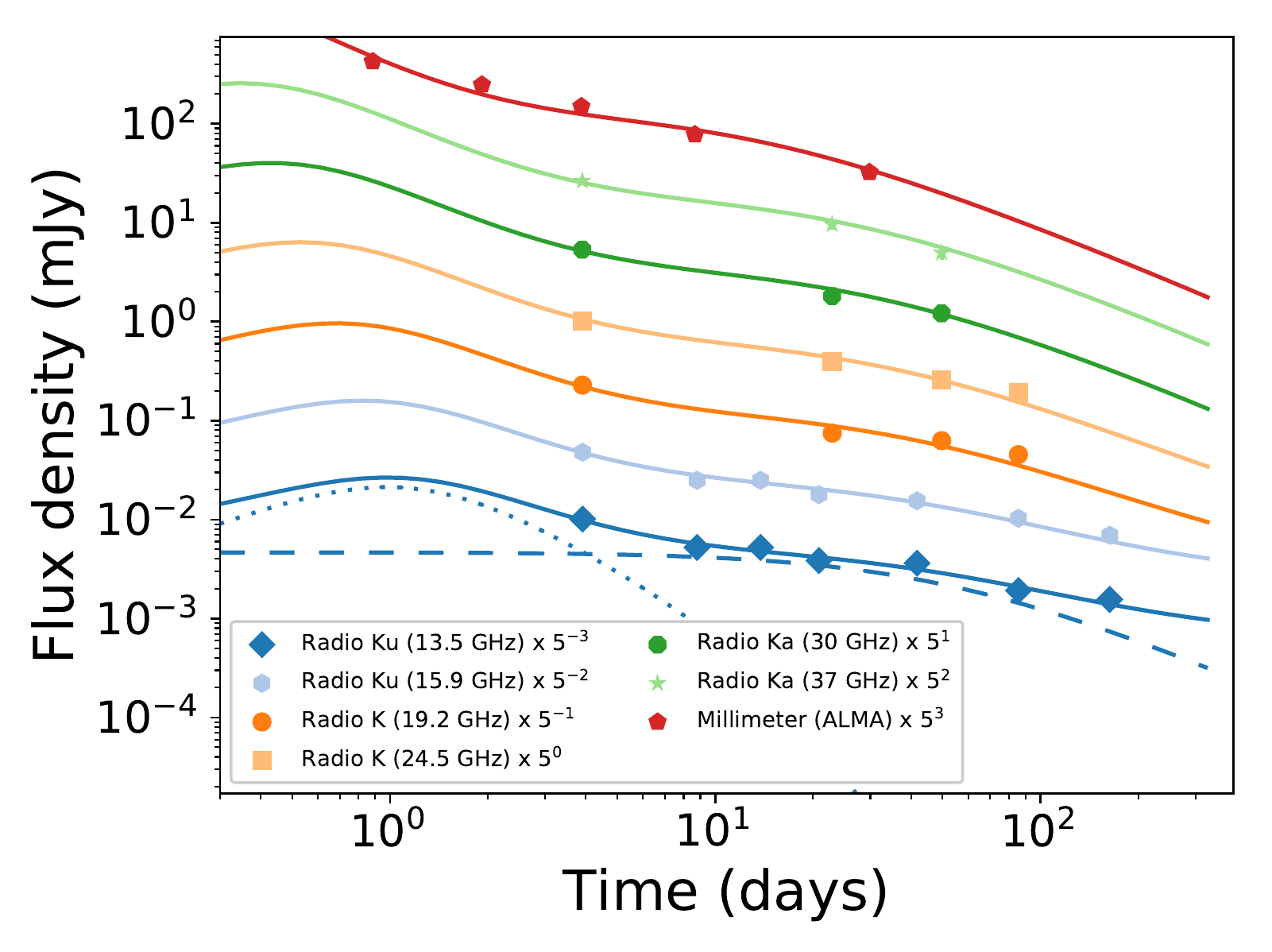} \\ [-3pt]
  \includegraphics[width=0.47\textwidth]{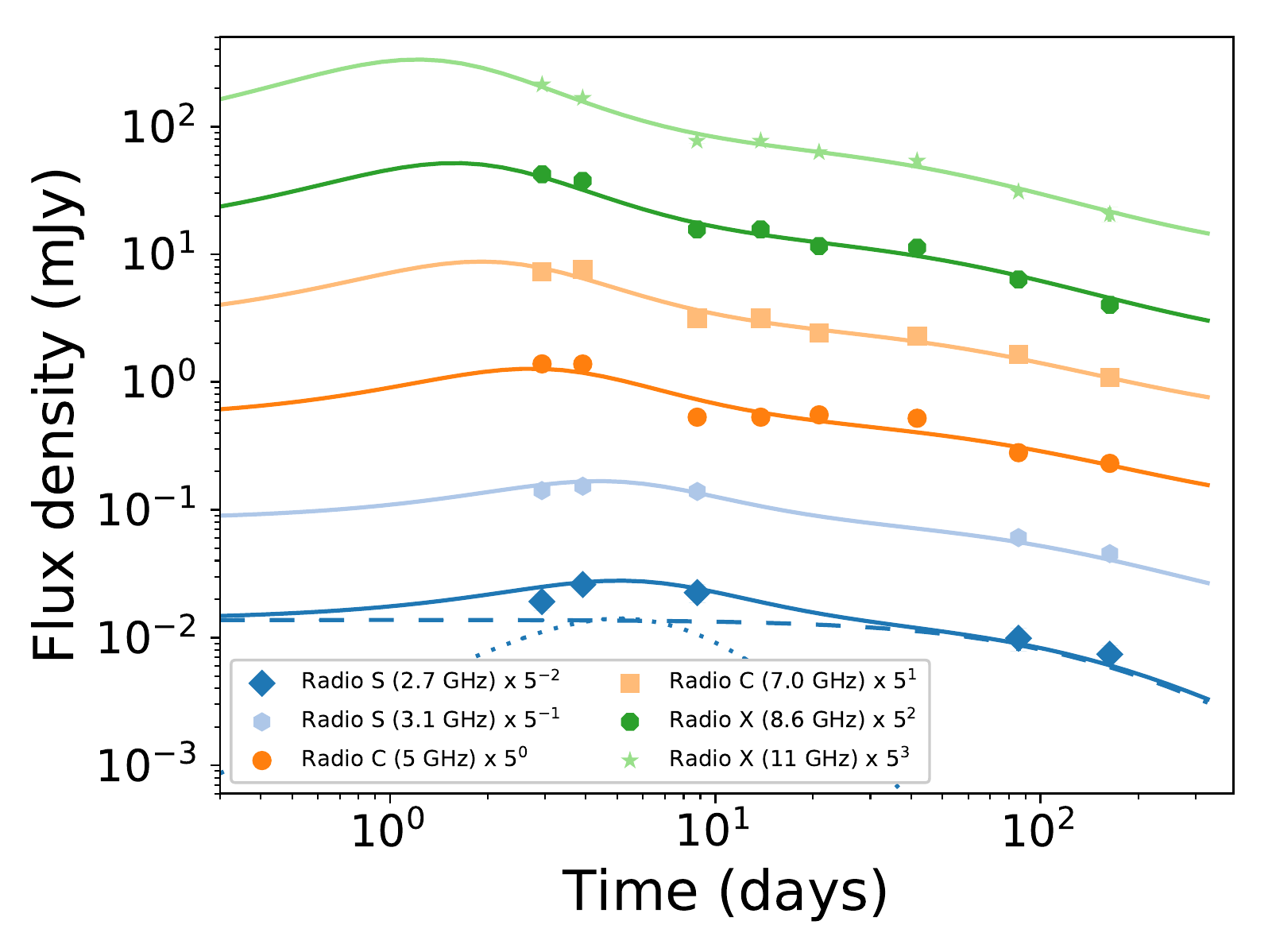} &
  \includegraphics[width=0.47\textwidth]{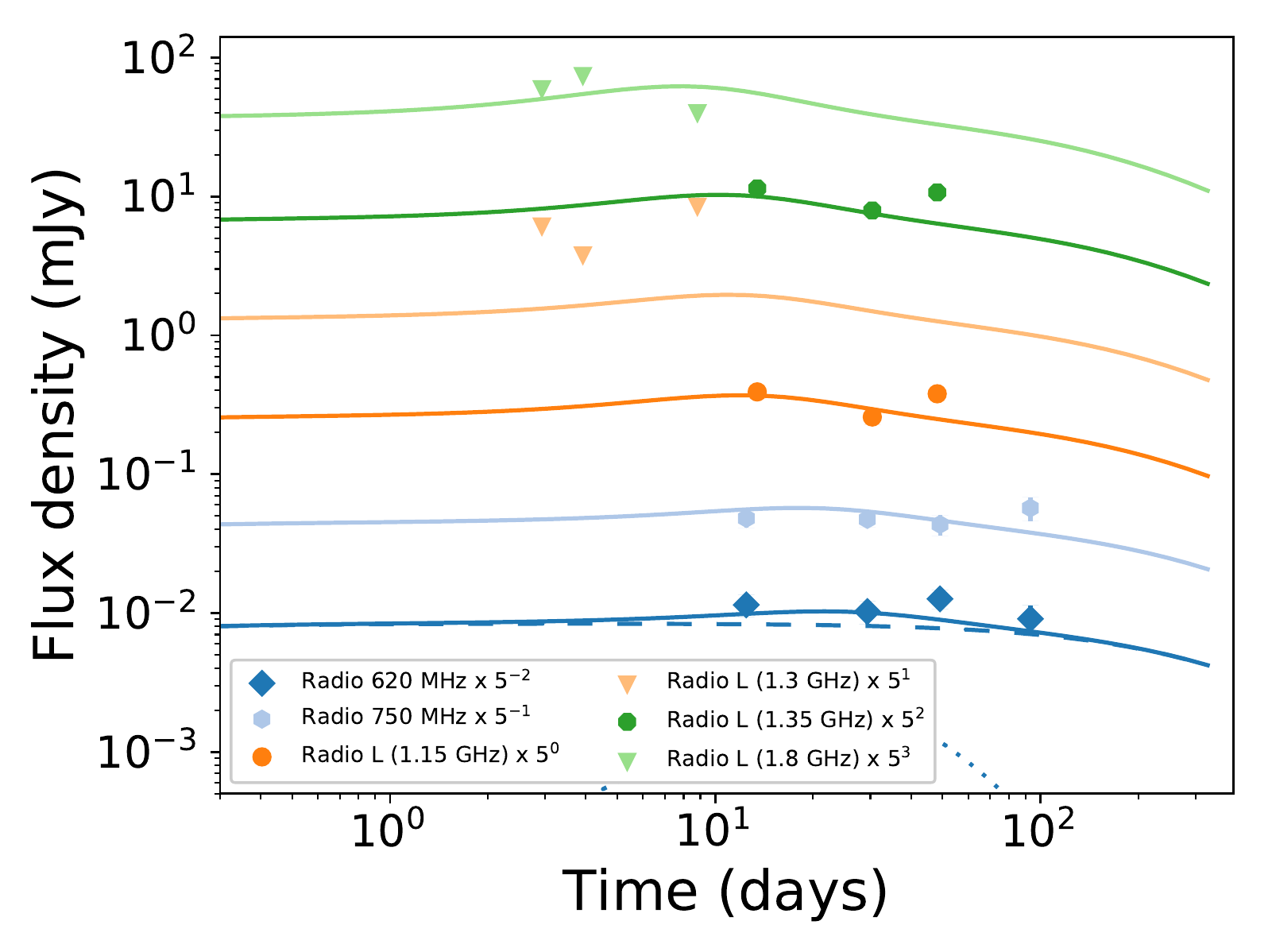} \\ [-3pt]
 \end{tabular}
 \caption{X-ray (top left), UV/optical (top right), optical/NIR (center left), and radio 
(center right and bottom) light curves of GRB\,181201A, together with a full FS+RS model (solid 
lines). Data represented by open symbols are not included in the model fit.
We show a decomposition of the X-ray, \Swift/{\textit w2}-band, optical $R$-band, 
13.5\,GHz, 2.7~GHz, and 620\,MHz light curves into FS (dashed) and RS (dotted) components.
Light curves exhibiting a late-time flattening incorporate a contribution from the underlying host 
galaxy (Section~\ref{text:MCMC}). The combined model explains the overall behavior of the light 
curves at all 36 observing frequencies. See Fig.~\ref{fig:modellc_RS} for a 
combined plot showing all lightcurves together. 
}
\label{fig:modellc_RS_splits}
\end{figure*}

\begin{figure*}
\begin{tabular}{ccc}
 \centering
 \includegraphics[width=0.31\textwidth]{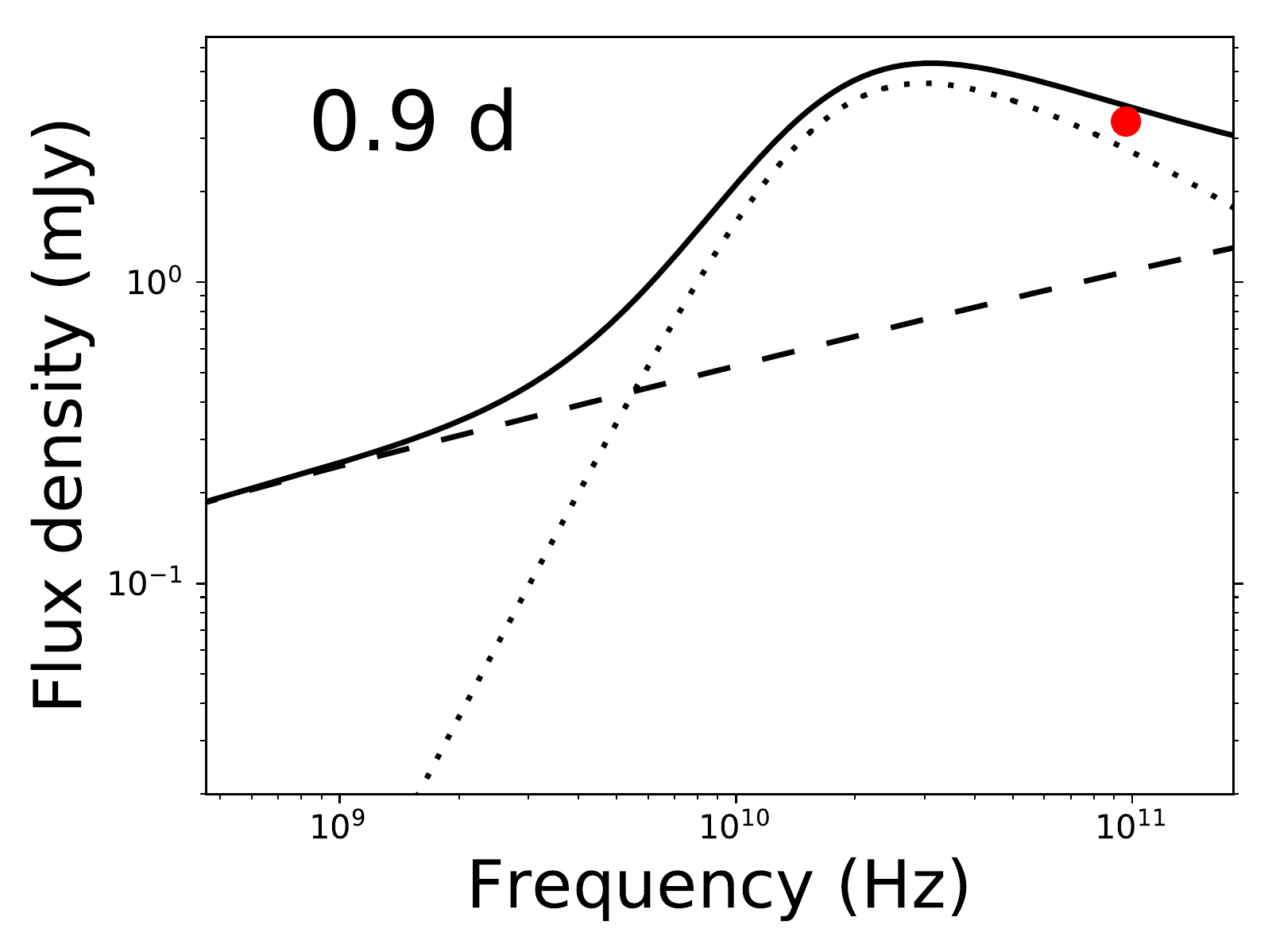} &
 \includegraphics[width=0.31\textwidth]{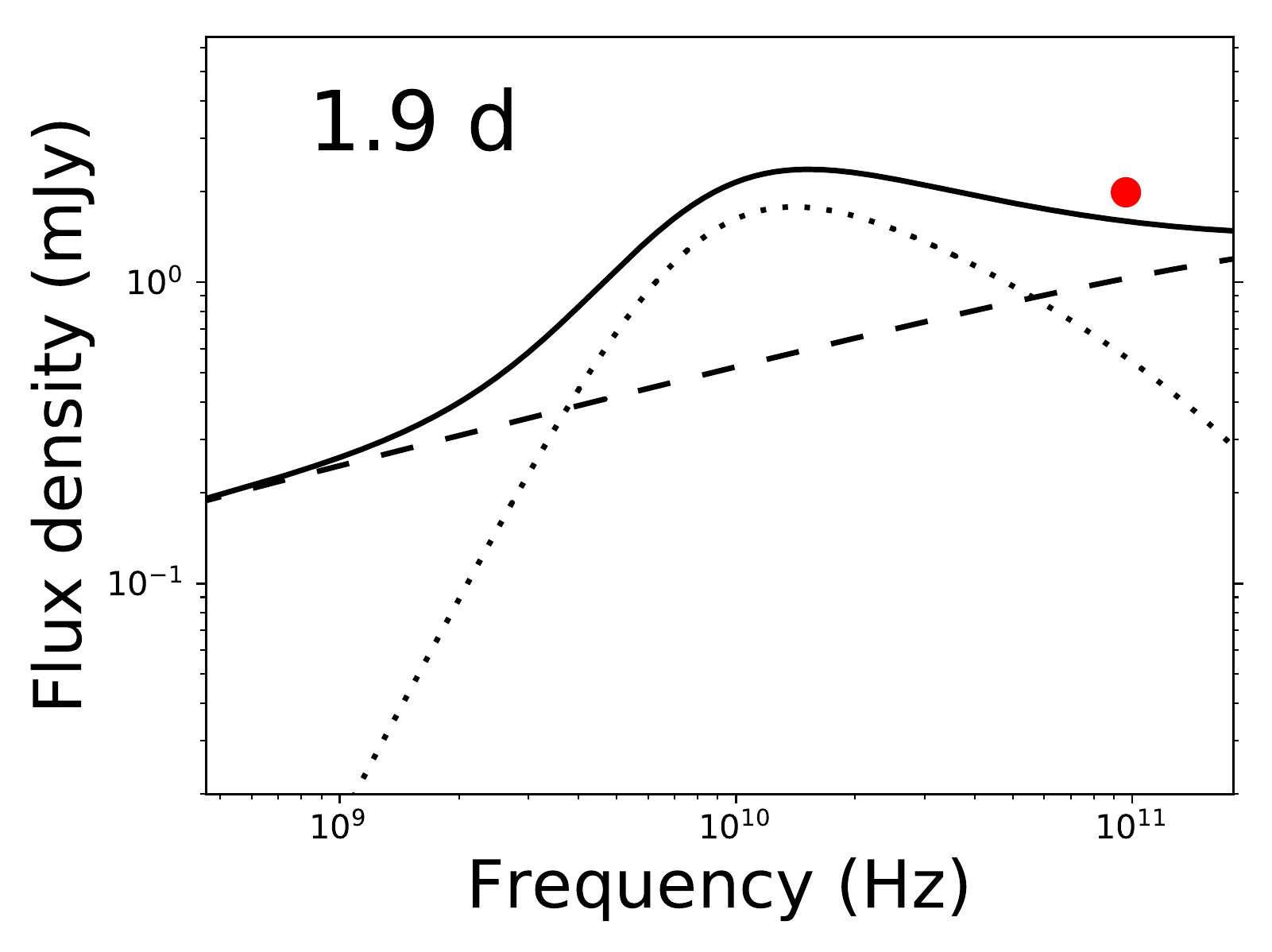} &
 \includegraphics[width=0.31\textwidth]{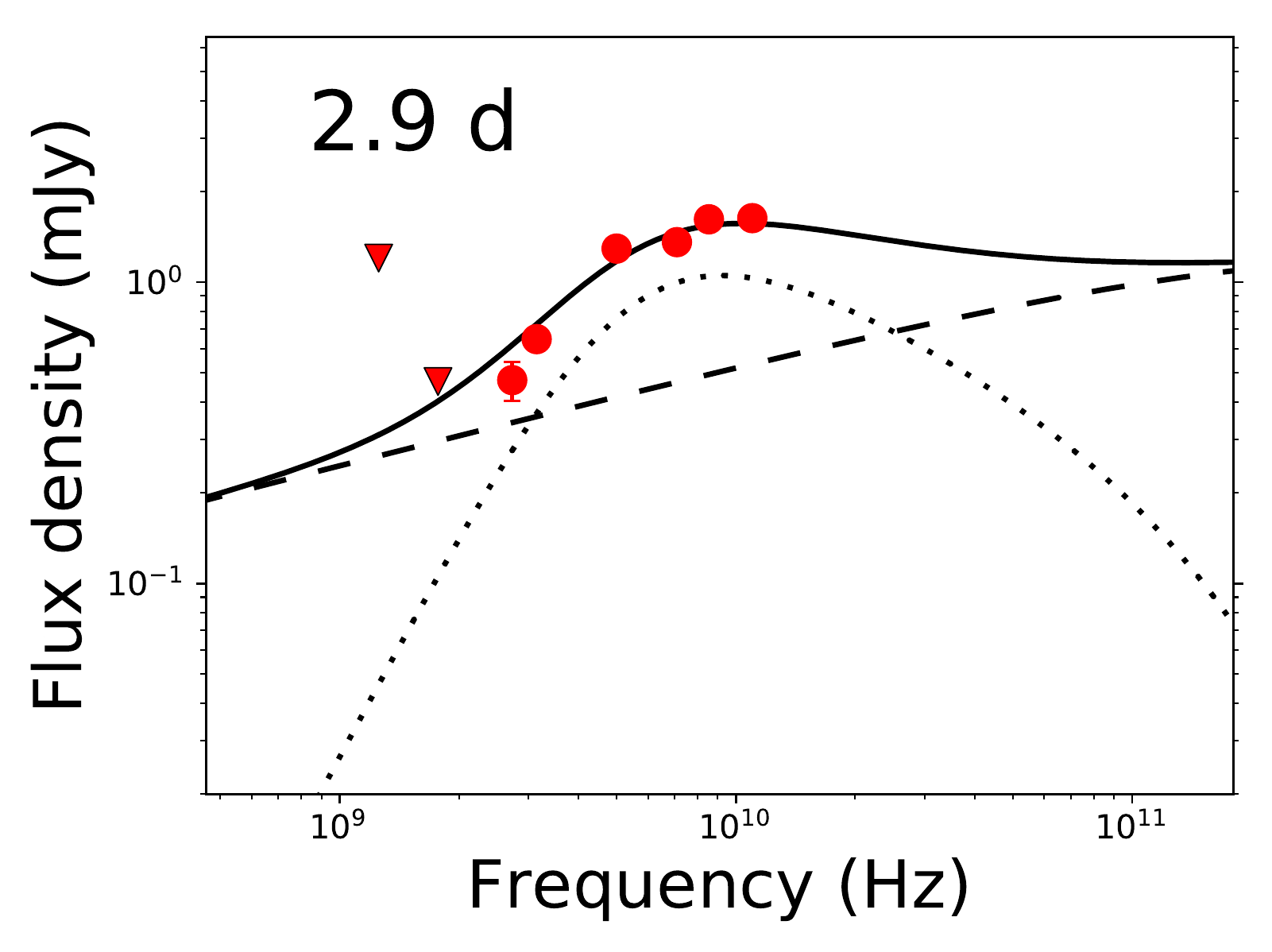} \\
 \includegraphics[width=0.31\textwidth]{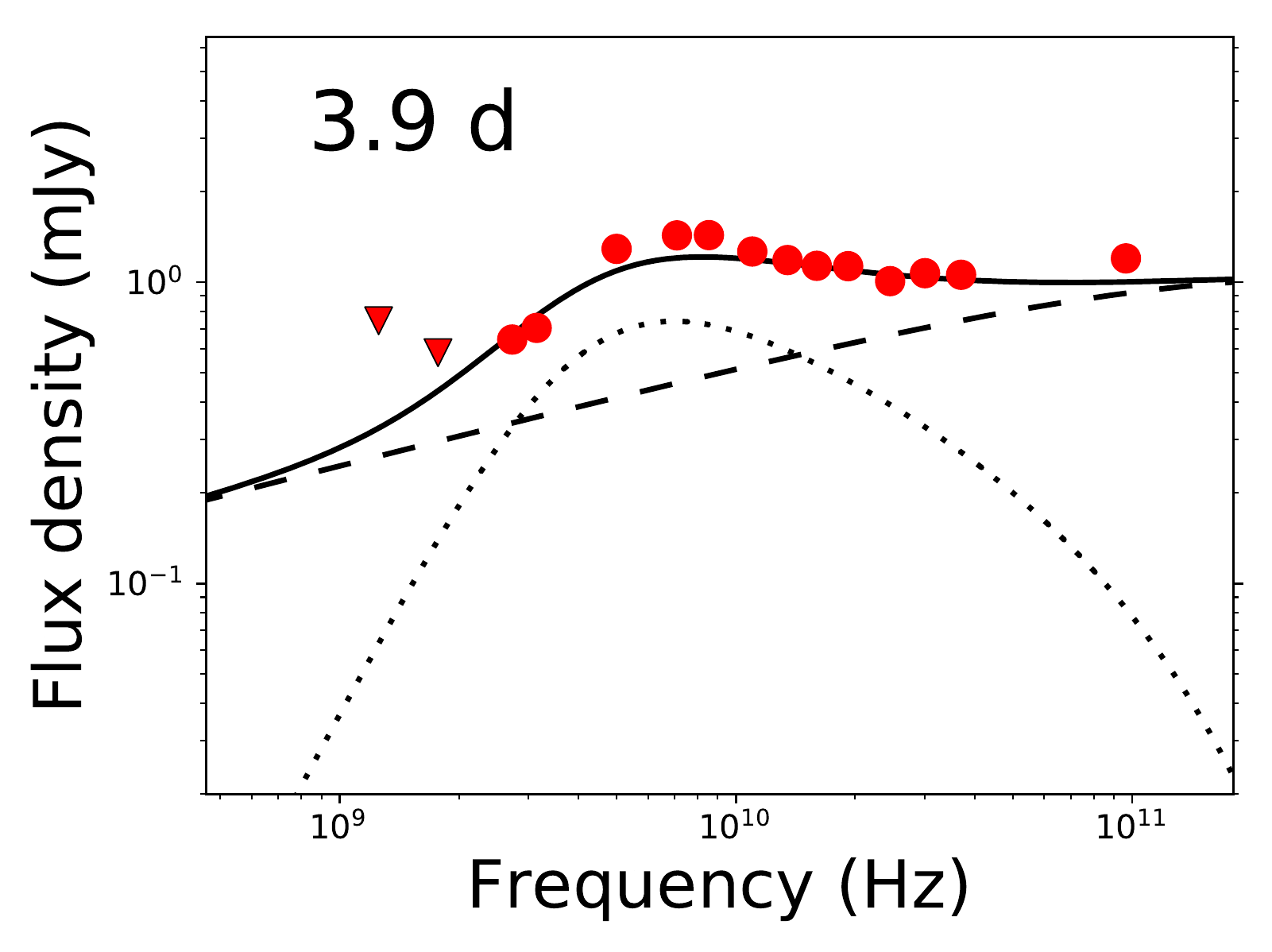} &
 \includegraphics[width=0.31\textwidth]{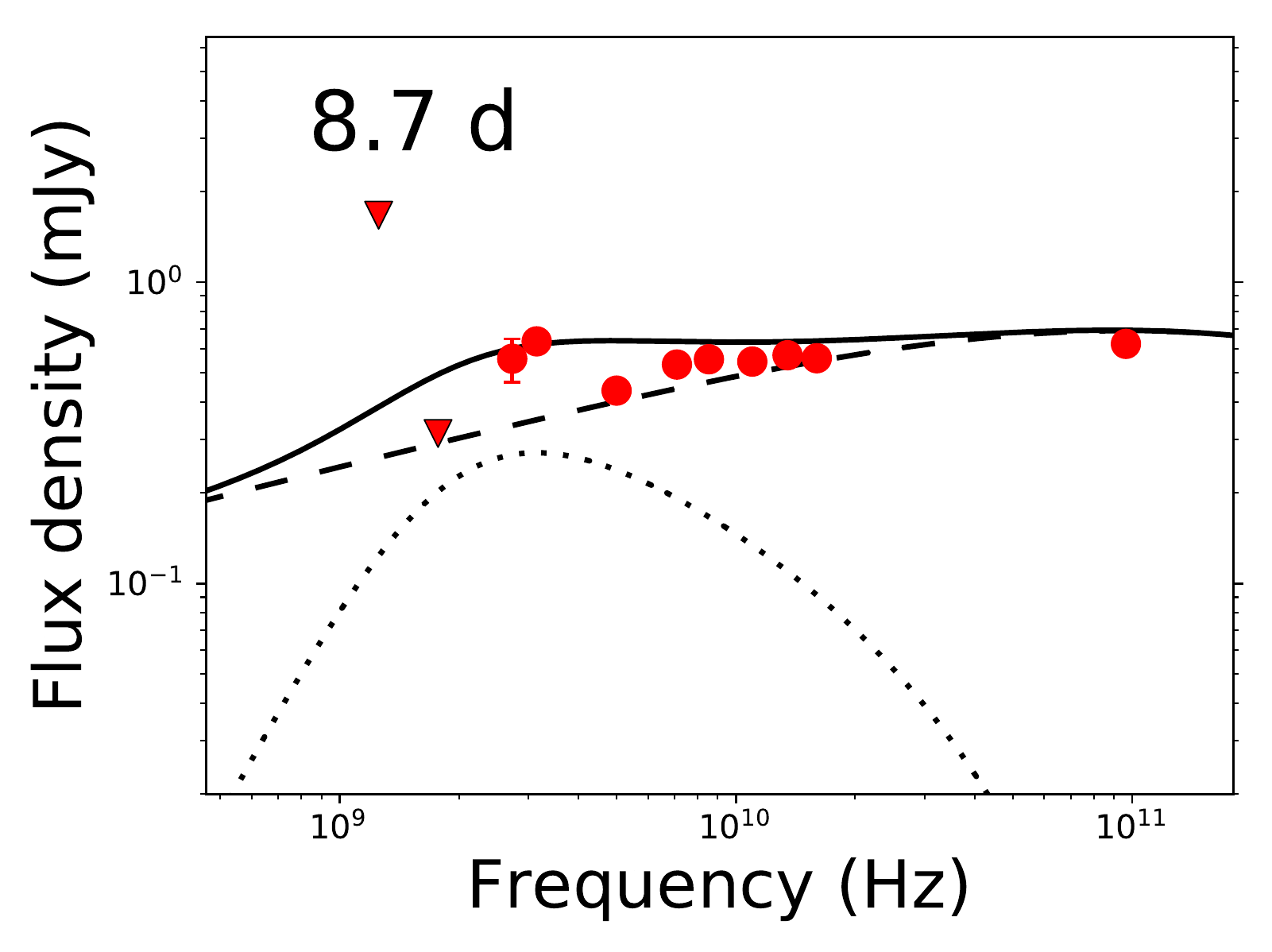} &
 \includegraphics[width=0.31\textwidth]{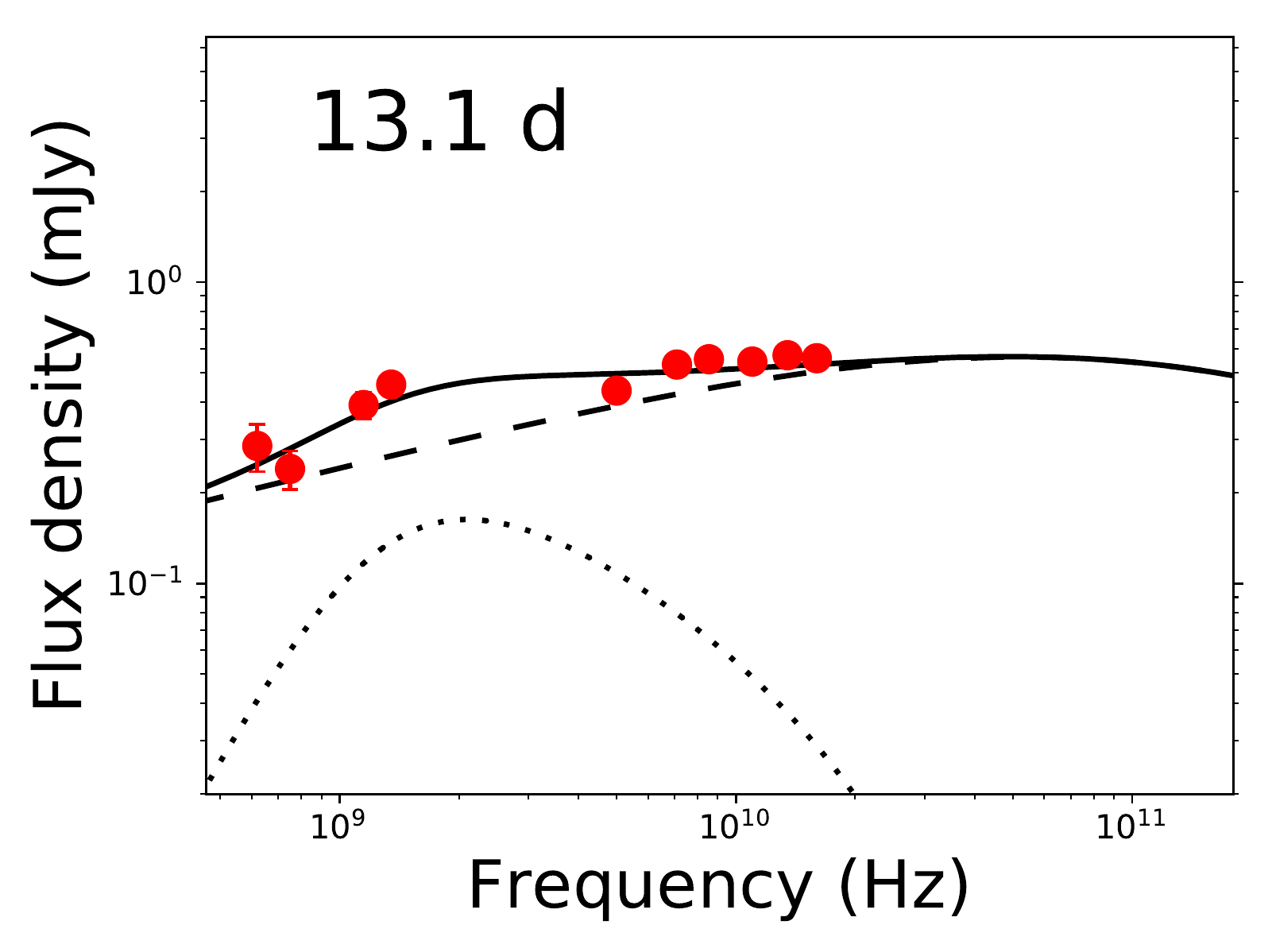} \\
 \includegraphics[width=0.31\textwidth]{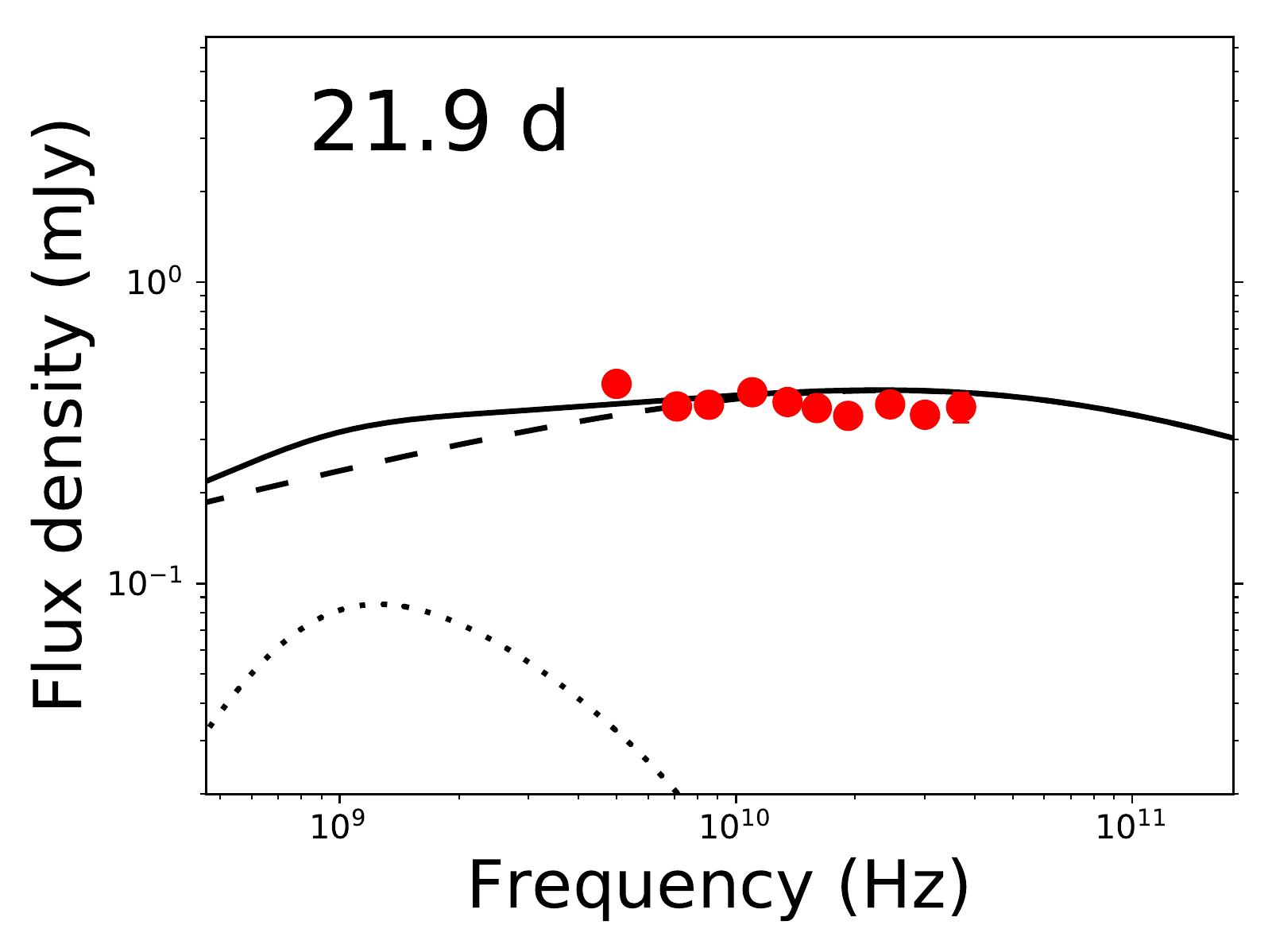} &
 \includegraphics[width=0.31\textwidth]{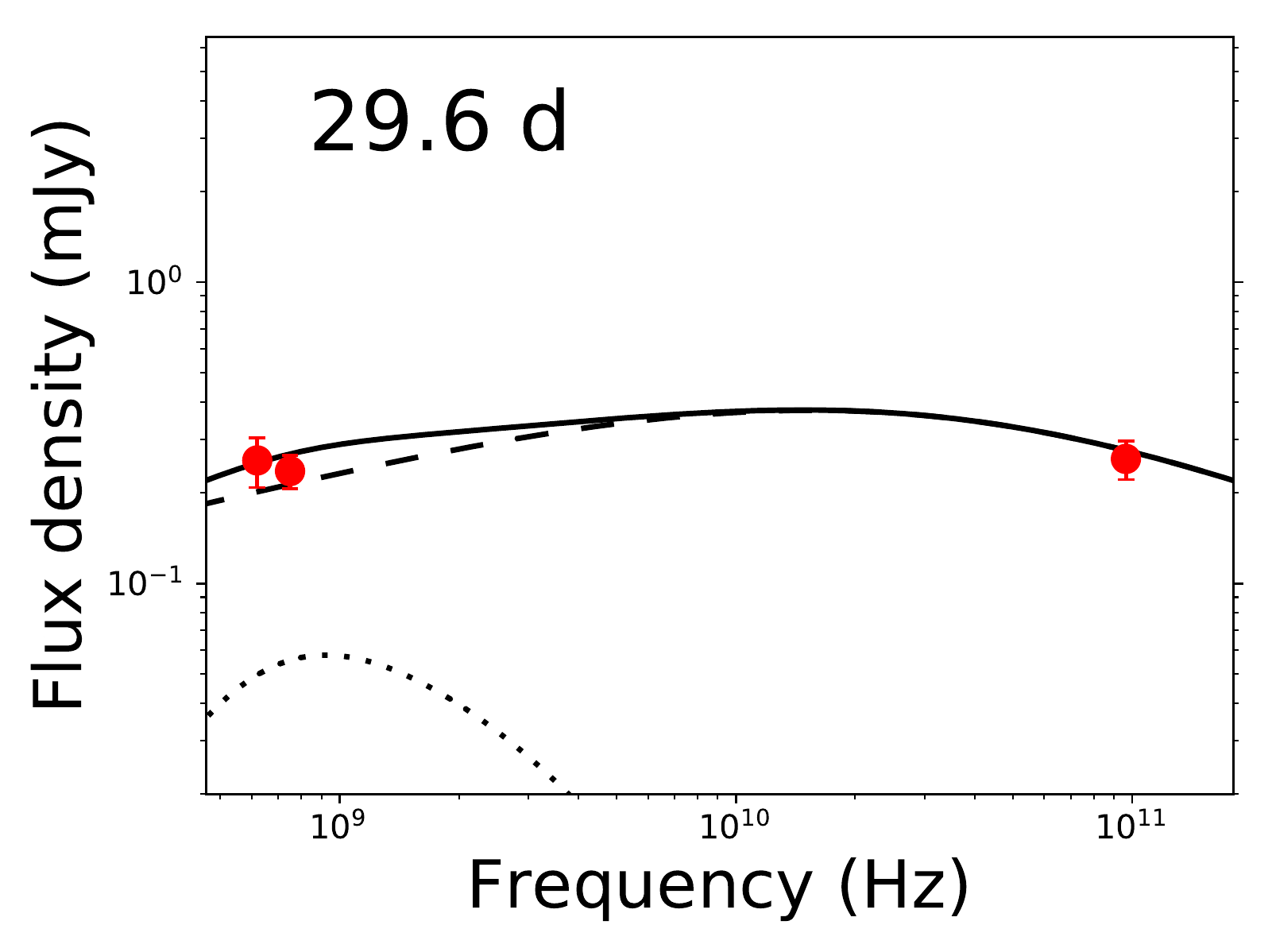} &
 \includegraphics[width=0.31\textwidth]{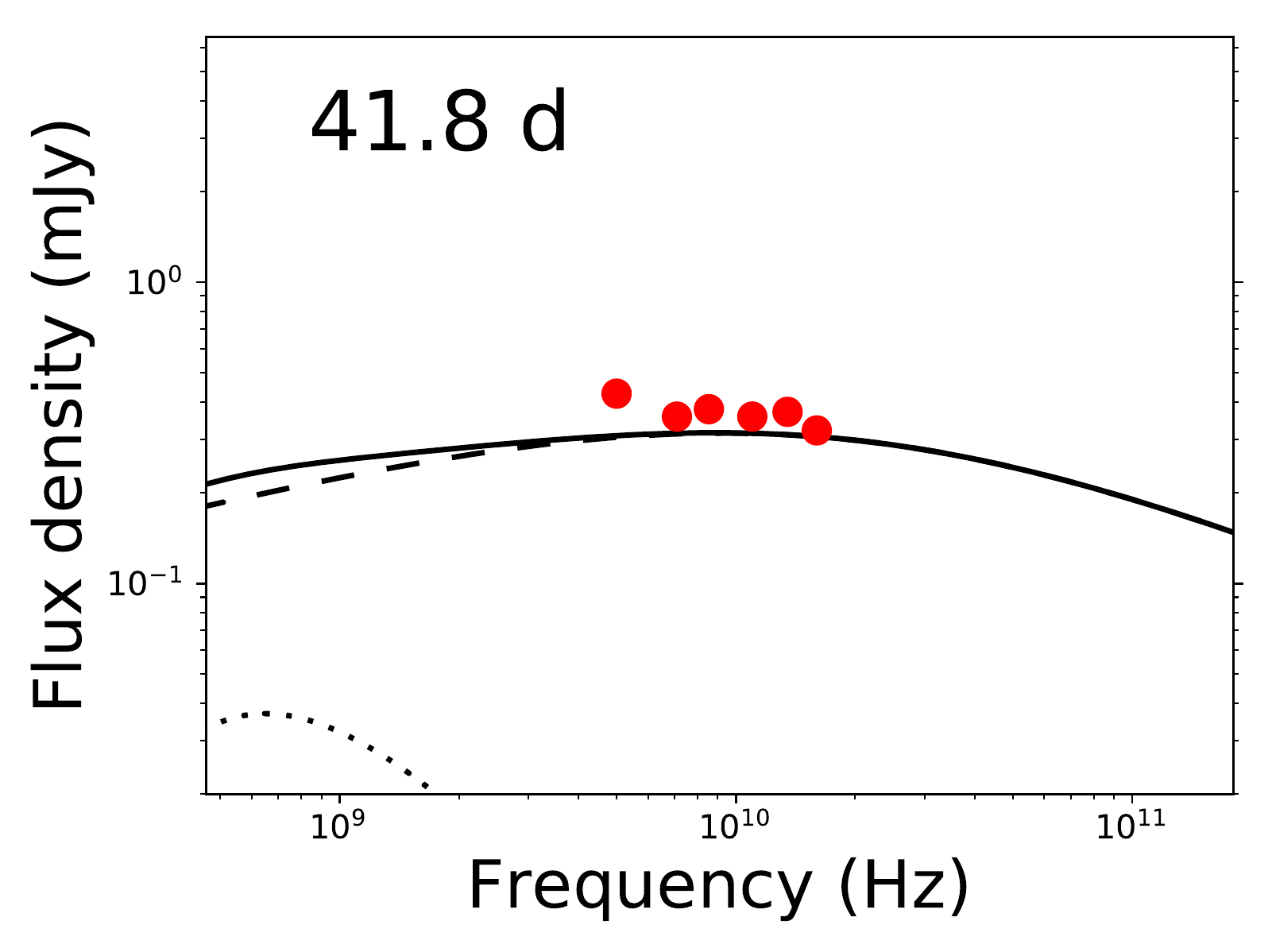} \\
 \includegraphics[width=0.31\textwidth]{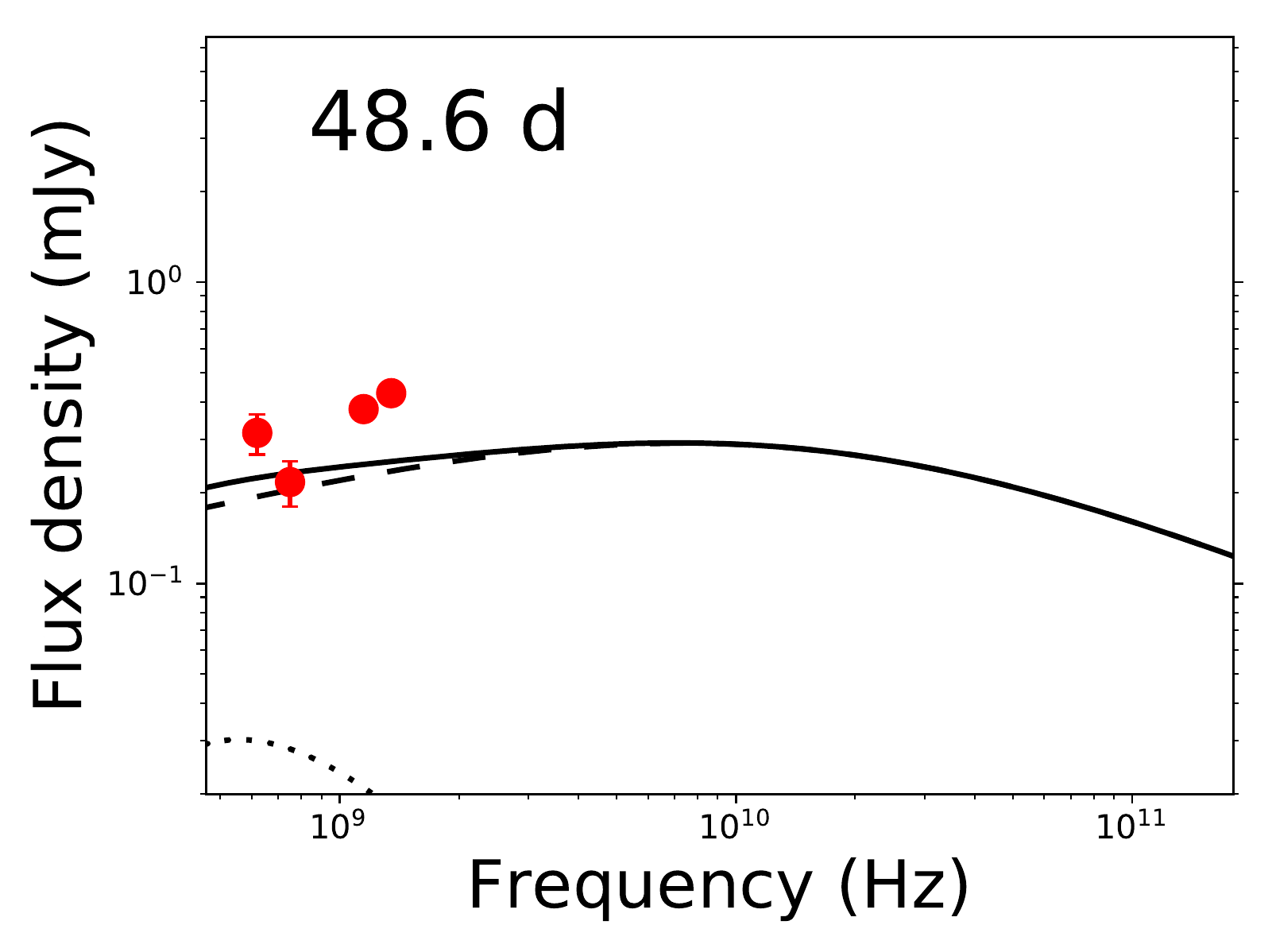} & 
 \includegraphics[width=0.31\textwidth]{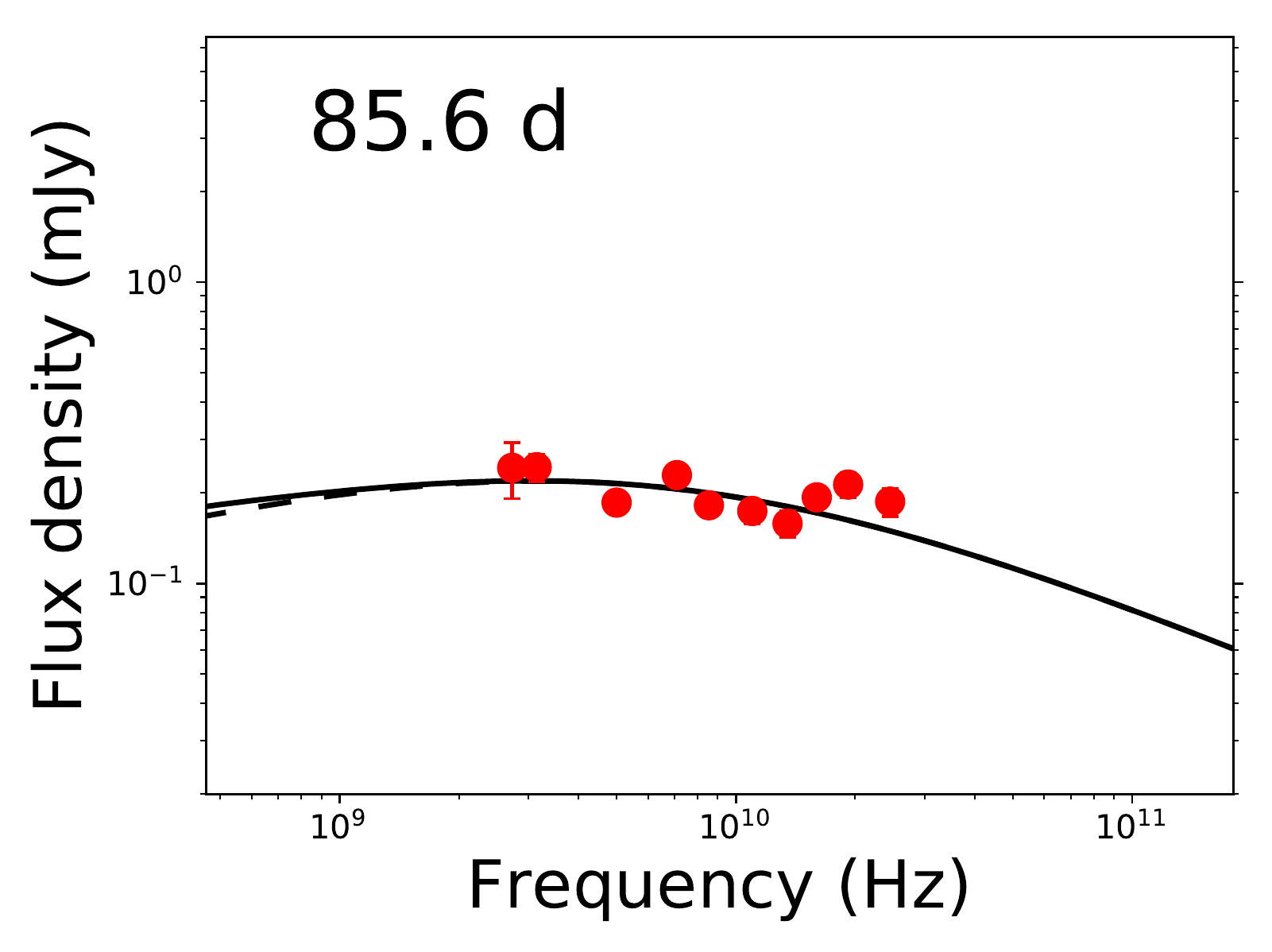} & 
 \includegraphics[width=0.31\textwidth]{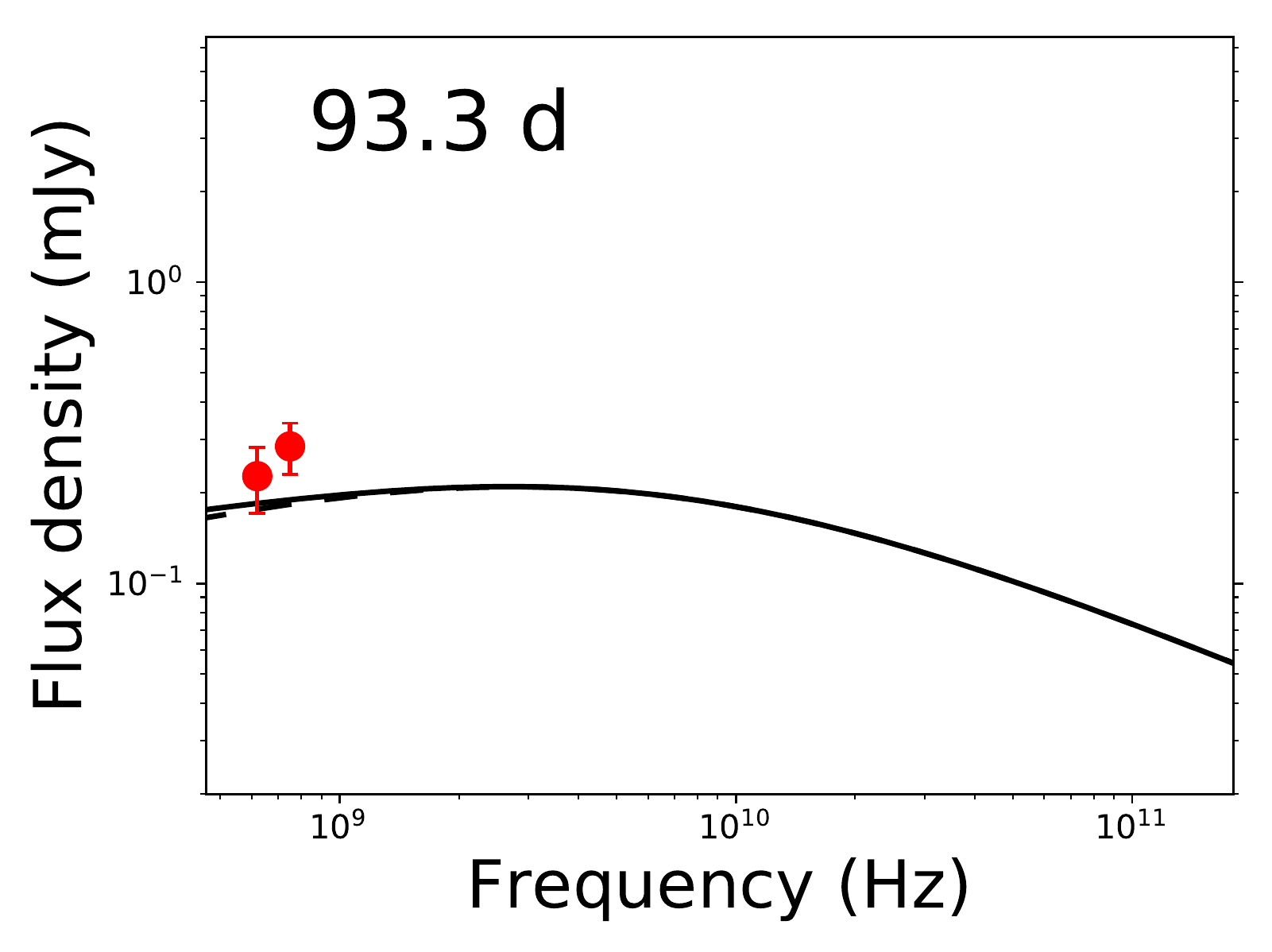} \\
 \includegraphics[width=0.31\textwidth]{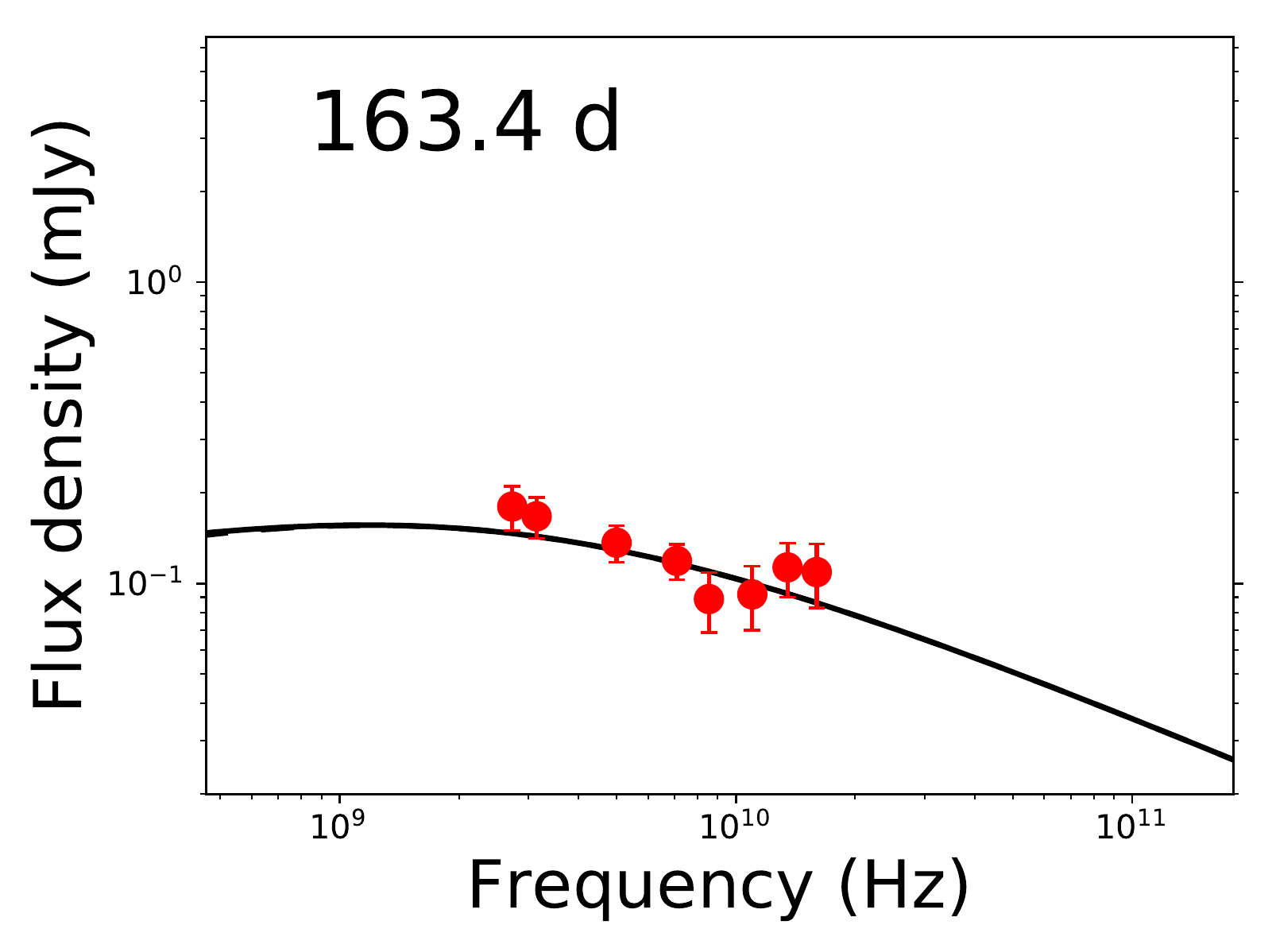} & & \\
\end{tabular}
\caption{
VLA cm-band and ALMA mm-band spectral energy distributions of GRB~181201A at multiple epochs 
starting at 0.9~days, together with the same FS+RS wind model as in Figure 
\ref{fig:modellc_RS_splits} (solid), decomposed into FS (dashed) and RS (dotted) contributions. The 
contribution of the host galaxy (included in the modeling) has been subtracted from the measured 
flux density at each frequency. See Fig.~\ref{fig:modelseds} for a plot 
showing these SEDs together with optical and X-ray observations.}
\label{fig:modelsed_RS}
\end{figure*}

\begin{figure} 
  \includegraphics[width=\columnwidth]{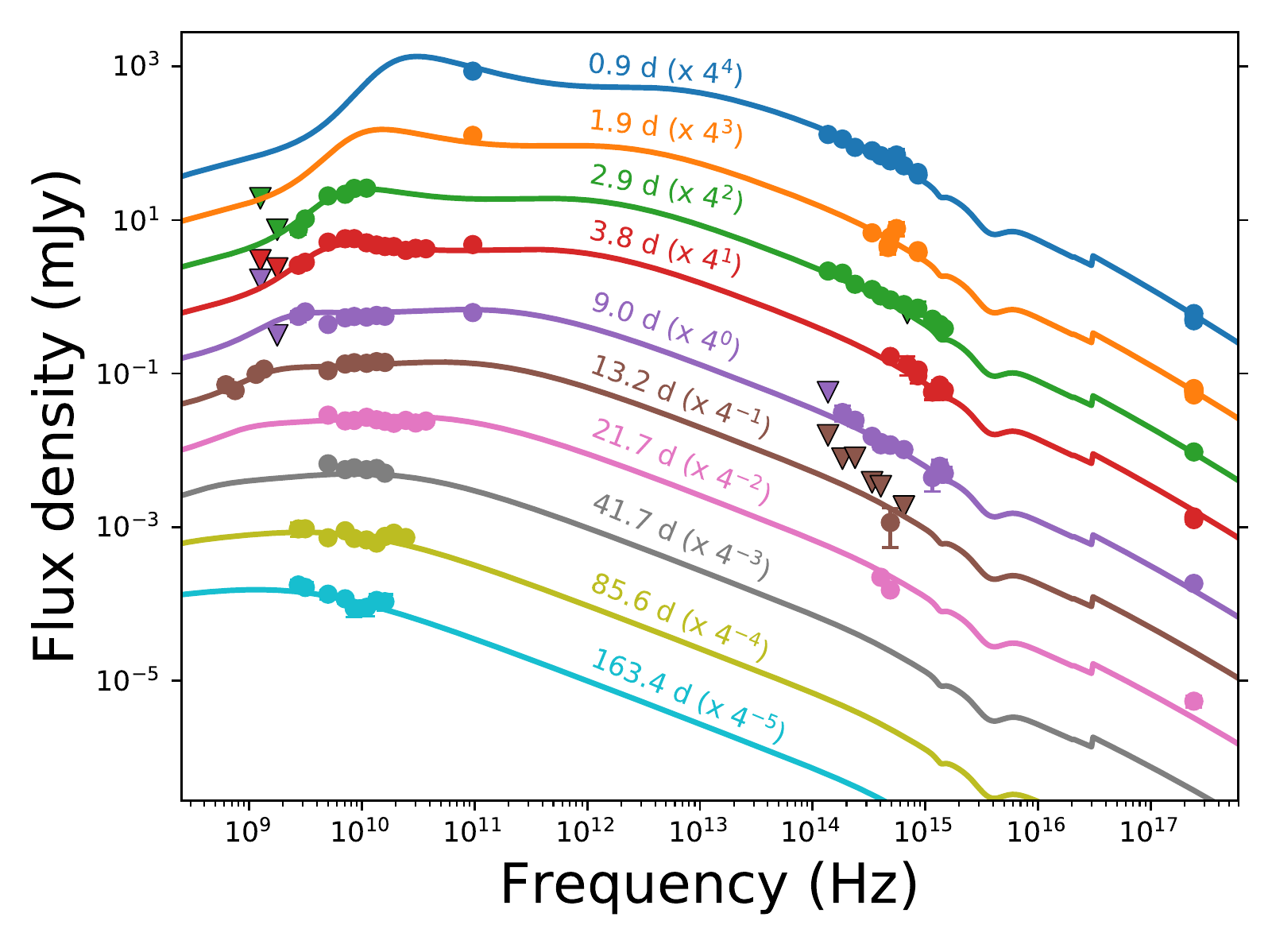}
 \caption{Multi-band SEDs of GRB\,181201A from $\approx0.9$ to $\approx163.4$~days, together with 
the FS+RS model presented in Section \ref{text:multimodel}. The combined model explains the overall 
behavior of the afterglow spanning over 8 orders of magnitude in frequency and over 2 orders of 
magnitude in time.}
\label{fig:modelseds}
\end{figure}

\begin{figure*} 
  \includegraphics[width=\textwidth]{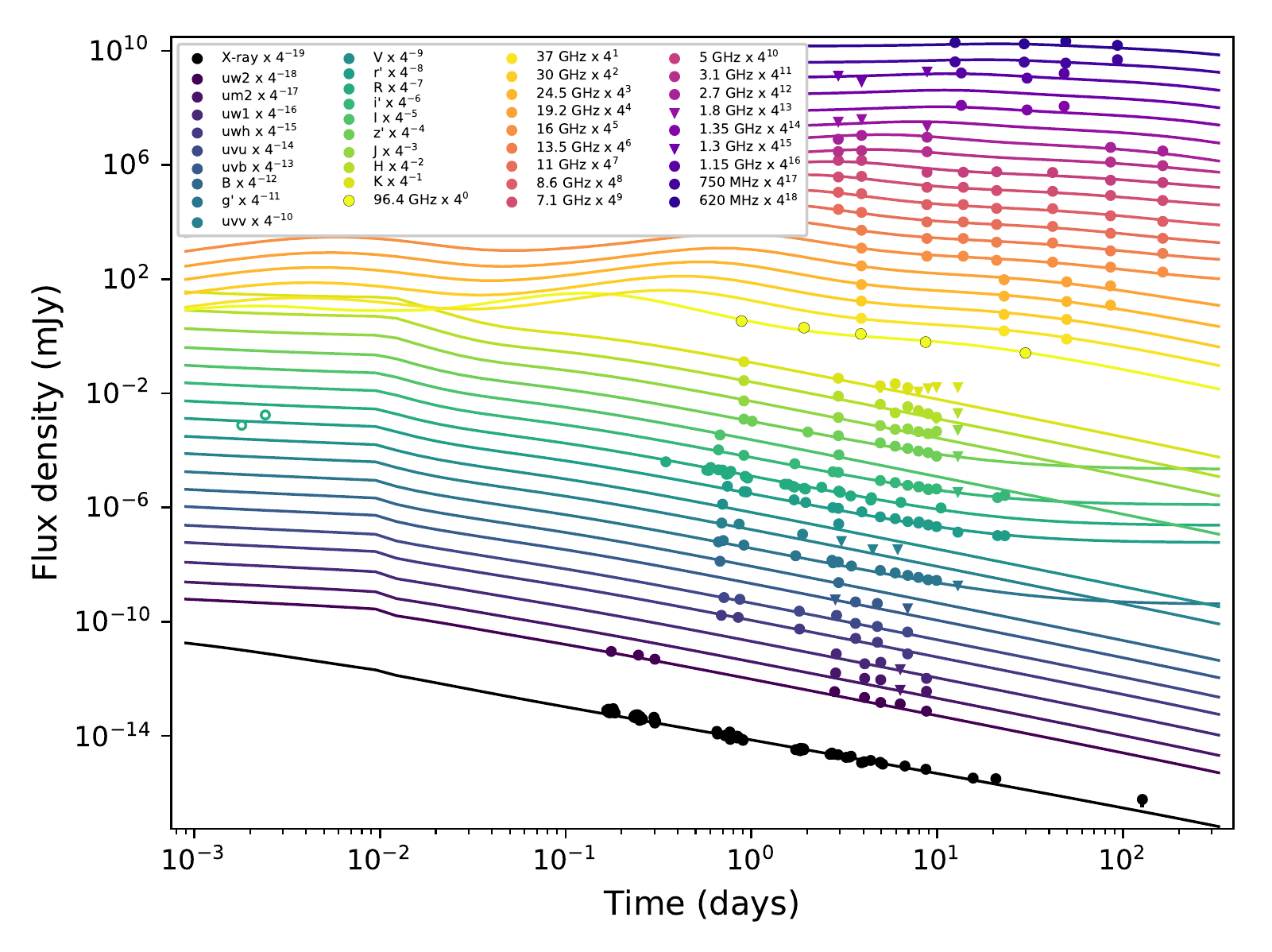}
 \caption{X-ray to radio light curves of GRB\,181201A, together with the FS+RS model presented in 
Section \ref{text:multimodel}, presented together for reference. The combined model explains the 
overall behavior of the light curves at all 36 observing frequencies.}
\label{fig:modellc_RS}
\end{figure*}

\begin{figure*}
  \centering
  \includegraphics[width=\textwidth]{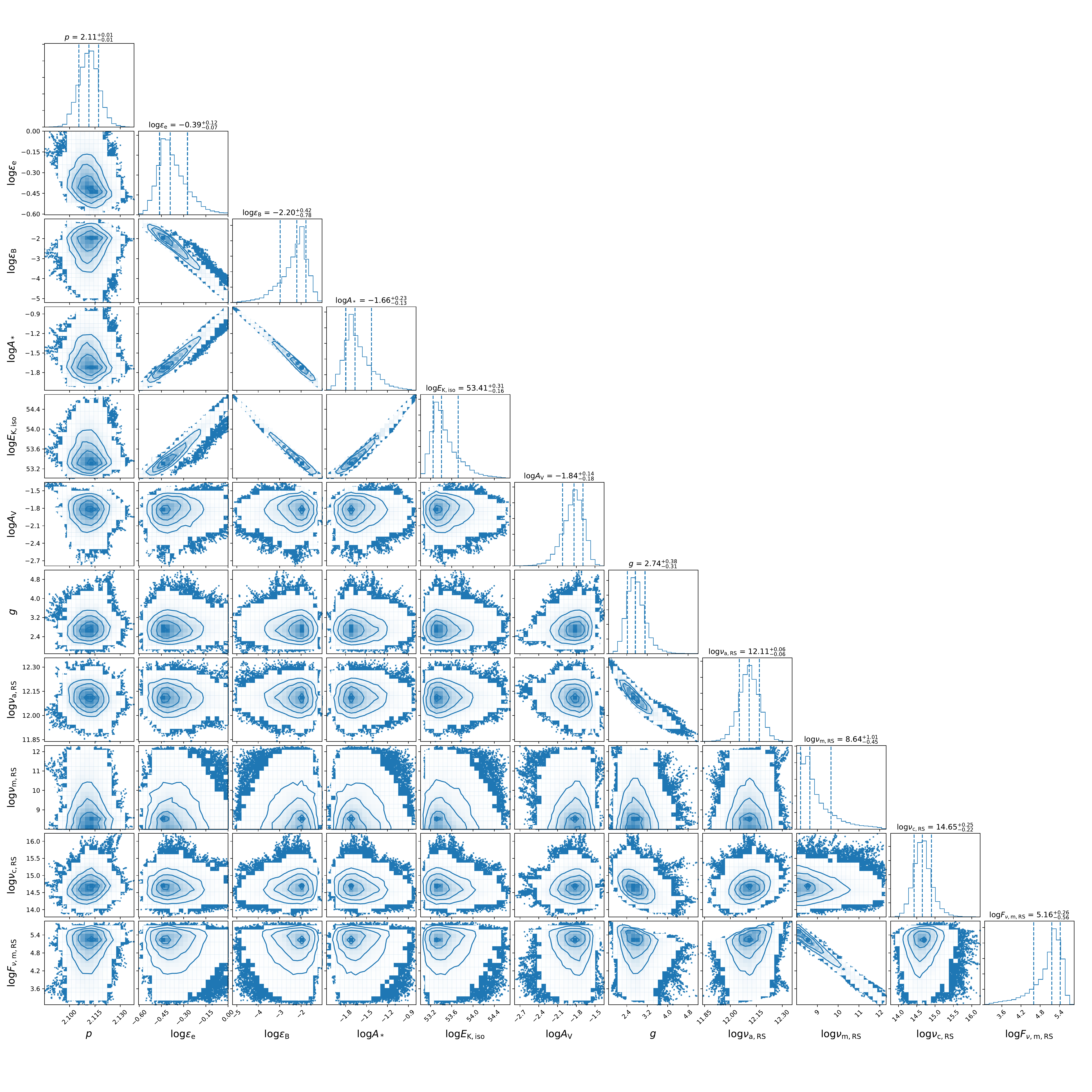}
  \caption{Correlations and marginalized posterior density for all the free parameters in 
the RS+FS model (not including the host galaxy fluxes). Frequencies are in units of Hz, $\fnumr$ in 
mJy, and $E_{\rm K,iso}$ in erg. 
The contours enclose 39.3\%, 86.4\% and 
98.9\% of the probability mass in each correlation plot (corresponding to $1\sigma$, $2\sigma$, 
and $3\sigma$ regions for two-dimensional Gaussian distributions), while the dashed lines in the 
histograms indicate the 15.9\%, 50\% (median), and 84.1\% quantiles (corresponding to $\pm1\sigma$ 
for one-dimensional Gaussian distributions).}
\label{fig:corner}
\end{figure*}

\section{Multi-wavelength Modeling}
\label{text:multimodel}
\subsection{Markov Chain Monte Carlo Analysis}
\label{text:MCMC}
Guided by the basic considerations discussed above, we fit all available X-ray to radio 
observations with a model comprising an FS and RS produced by a relativistic jet propagating in a 
wind-like circum-burst environment. The free parameters for the FS model are $p$, $E_{\rm 
K,iso,52}$, $A_*$, $\epse$, $\epsb$, and the extinction, $\AV$. Given the relatively shallow 
decline rate of the mm-band light curve ($\alpha\approx-0.72$) corresponding to the RS component, 
we focus on the non-relativistic RS model, which allows for shallower light curve decay than a 
relativistic RS. 
The free parameters for the RS model\footnote{We opt to fit for the observable quantities rather 
than the derived quantities of $\Gammajet$, $\tdec$, and $\RB$ for the RS, as these observables
uniquely specify the RS spectrum independent of the FS parameters during the fitting process.} 
are the break frequencies $\nuar$, $\numr$, $\nucr$ and the RS peak flux, $\fnumr$ all at a 
reference time, $t_{\rm ref}\equiv0.01$~days selected to be $\approx {\rm few}\times T_{90}$ (and 
hence comparable to $\tdec$). Additionally, we allow the parameter $g$, corresponding to the 
evolution of the RS Lorentz factor with radius after the deceleration time ($\Gamma\propto R^{-g}$) 
to vary \citep{mr99,ks00}. Finally, we allow for a constant contribution to the light curves from 
an underlying host galaxy in bands where late-time flattening is evident (optical $griz$ bands and 
the cm-bands from 2.7 to 24.5~GHz). All free parameters, including RS and FS components and the 
underlying host galaxy contamination, are fit simultaneously. Further details of the modeling 
procedure, including the likelihood function employed, are described elsewhere 
\citep{lbz+13,lbt+14,lab+16}. 

We performed a Markov Chain Monte Carlo (MCMC) exploration of the multi-dimensional parameter space 
using \texttt{emcee} \citep{fhlg13}. We ran 512 walkers for 10k steps and discarded a few 
($\lesssim0.15\%$) initial non-converged steps with low log likelihood as burn-in. Our highest 
likelihood model with
$p\approx2.1$, 
$\epse\approx0.37$
$\epsb\approx9.6\times10^{-3}$,
$E_{\rm K,iso}\approx2.2\times10^{53}$~erg,
$\Astar\approx1.9\times10^{-2}$,
and negligible extinction, $\AV\lesssim0.02$~mag,
together with an RS model with 
$g\approx2.7$, 
$\nuar\approx4.2$~GHz,
$\nucr\approx90$~GHz, and
$F_{\nu,\rm a,r}\approx0.6$~mJy
at $3.9$~days 
fits the X-ray to radio light curves and radio SEDs 
well (Figs.~\ref{fig:modellc_RS_splits} and \ref{fig:modelsed_RS}). 
The peak in the cm-band spectrum at 3.9~days corresponds to $\nuar$. 
The sum of the RS self-absorbed spectrum and the FS optically thin $\nu^{1/3}$ spectrum below 
$F_{\nu,\rm a,r}$ (Fig.~\ref{fig:modelsed_RS}) explains the intermediate $\beta\approx1.4$ 
spectral index below the cm-band peak (Section~\ref{text:4dsed}).
We present a selection of SEDs spanning radio to X-rays in Figure~\ref{fig:modelseds}, 
and all light curves together in Figure~\ref{fig:modellc_RS}.

For this model, the FS cooling frequency, $\nucf\approx2.6\times10^{16}$~Hz at 1\,d lies 
between the optical and X-rays, as expected from the basic considerations in 
Section~\ref{text:densityprofile}. The proximity of $\nucf$ to the X-ray band 
($\nucf\approx0.3$~keV at 8~days) explains the moderately steep decline in the X-ray light 
curve, compared to the expectation for $\nux\gtrsim\nucf$. 
Furthermore, $\nuaf\approx60$~MHz is below the VLA and GMRT frequencies, and is therefore not 
constrained, resulting in degeneracies between the model parameters (Fig.~\ref{fig:corner}). The FS 
characteristic frequency at 4~days, $\numf\approx6\times10^{11}$~Hz is above the ALMA band, as 
expected from the radio/mm SED at that time (Section~\ref{text:4dsed}). We list the properties of 
the best fit model and summarize the marginalized posterior density function in 
Table~\ref{tab:params}.

\subsection{Determination of Ejecta Initial Lorentz Factor}
\label{text:Gammajet}
We now use the MCMC results to derive the initial Lorentz factor of the jet. The jet's initial 
Lorentz factor can be determined from joint RS and FS observations by solving the following 
equations self-consistently \citep{zkm03,lab+18} at the deceleration time, $\tdec$,
\begin{align}
 \frac{\numr}{\numf} &\sim \Gammajet^{-2}\RB, \nonumber \\
 \frac\nucr\nucf &\sim \RB^{-3}, \nonumber \\
 \frac\fnumr\fnumf &\sim \Gammajet\RB,
 \label{eq:consistency}
\end{align}
where $\Gammajet$ is the initial Lorentz factor of the GRB jet, $\RB\equiv(\epsilon_{\rm 
B,r}/\epsilon_{\rm B,f})^{1/2}$ is the relative magnetization of the ejecta, and we have assumed 
similar values of $\epse$ and the Compton $y$ parameter in the two regions. The effective 
magnetizations of the shocked ejecta and shock external medium may well be very different, and the 
introduction of \RB\ allows us to investigate such variations caused, for instance, by the presence 
of magnetic fields produced in the initial GRB and advected in the outflow 
\citep{uso92,mr97,zkm03}.

This system of equations has three unknowns (\tdec, \RB, and $\Gamma_0$), and can be inverted 
exactly, to yield
\begin{align}
 \tdec &= t_{\rm ref} \left[
                        \left(\frac{\nucf^{\prime}}{\nucr^{\prime}}\right)
                        \left(\frac{\fnumf^{\prime}}{\fnumr^{\prime}}\right)^2
                        \left(\frac{\numf^{\prime}}{\numr^{\prime}}\right)
                        \right]^\tau, \nonumber \\
\tau &\equiv \left[\frac{15}{2(4-k)}-\frac{8(g+3)}{7(2g+1)}\right]^{-1}
\label{eq:tdec}
\end{align} 
where the primed quantities refer to values at the reference time of $t_{\rm ref}=0.01$~days used in
the MCMC analysis. For our best-fit model, this yields $\tdec\approx5.3\times10^{-3}$~days or 
$\approx460$~s; the full MCMC analysis gives $\log(\tdec/{\rm days})=-2.27^{+0.13}_{-0.14}$. We 
note that this is longer than the burst duration, $T_{90}\approx172$~s 
(Section~\ref{text:obs_basic}), consistent with the thin shell case \citep{zkm03}.

Given this value of $\tdec$, which satisfies equation \ref{eq:consistency}, we can in principle 
compute the initial Lorentz factor and magnetization,
\begin{align}
\RB = \left(\frac\nucf\nucr\right)^{1/3} \nonumber\\
\Gammajet = \frac\fnumr\fnumf\RB^{-1}.
\label{eq:jetpropcalc}
\end{align}
However, we find that whereas $\nucr$ and $\nucf$ are reasonably well constrained, $\fnumr$ is 
completely degenerate with $\numr$. This is because the RS peak in the radio band corresponds to 
$\nuar$, and hence the data do not constrain $\numr$. In the observed regime of $\numr<\nuar$, the 
peak flux density of the RS component, 
$F_{\nu,\rm a,r}\propto \fnumr(\nuar/\numr)^{(1-p)/2}$ results in the degeneracy, 
$\fnumr\propto\numr^{(1-p)/2}\propto\numr^{-0.5}$ for $p\approx2$, which exactly explains the 
observed correlation between these parameters (Figure~\ref{fig:corner}). Our MCMC samples span 4 
orders of magnitude in $\numr$ and correspondingly about 2 orders of magnitude in $\fnumr$. The 
sampling is clipped at the lower end in $\numr$ by the prior on $\numr>10^8$~Hz (corresponding to 
$\fnumr\lesssim10^8$~mJy) and at the lower end in $\fnumr$ by the actual observed peak flux density 
of the RS component (corresponding to $\numr\lesssim10^{12}$~Hz). This degeneracy between $\numr$ 
and $\fnumr$ does not affect the calculation of $\tdec$ (equation \ref{eq:tdec}), because $\numr$ 
and $\fnumr$ enter in the combination $\fnumr\numr^2\propto\numr^{2-p}$, which is only weakly 
dependent on $\numr$ for $p\approx2$ (i.e., the degenerate terms cancel out). Therefore, we can 
constrain $\tdec$ and $\RB$ through this system of equations, but not $\Gammajet$. We find 
$\RB\approx0.65$ for the best-fit solution, and $\RB=0.64^{+0.07}_{-0.08}$ from the MCMC.

\begin{figure}
  \centering
  \includegraphics[width=\columnwidth]{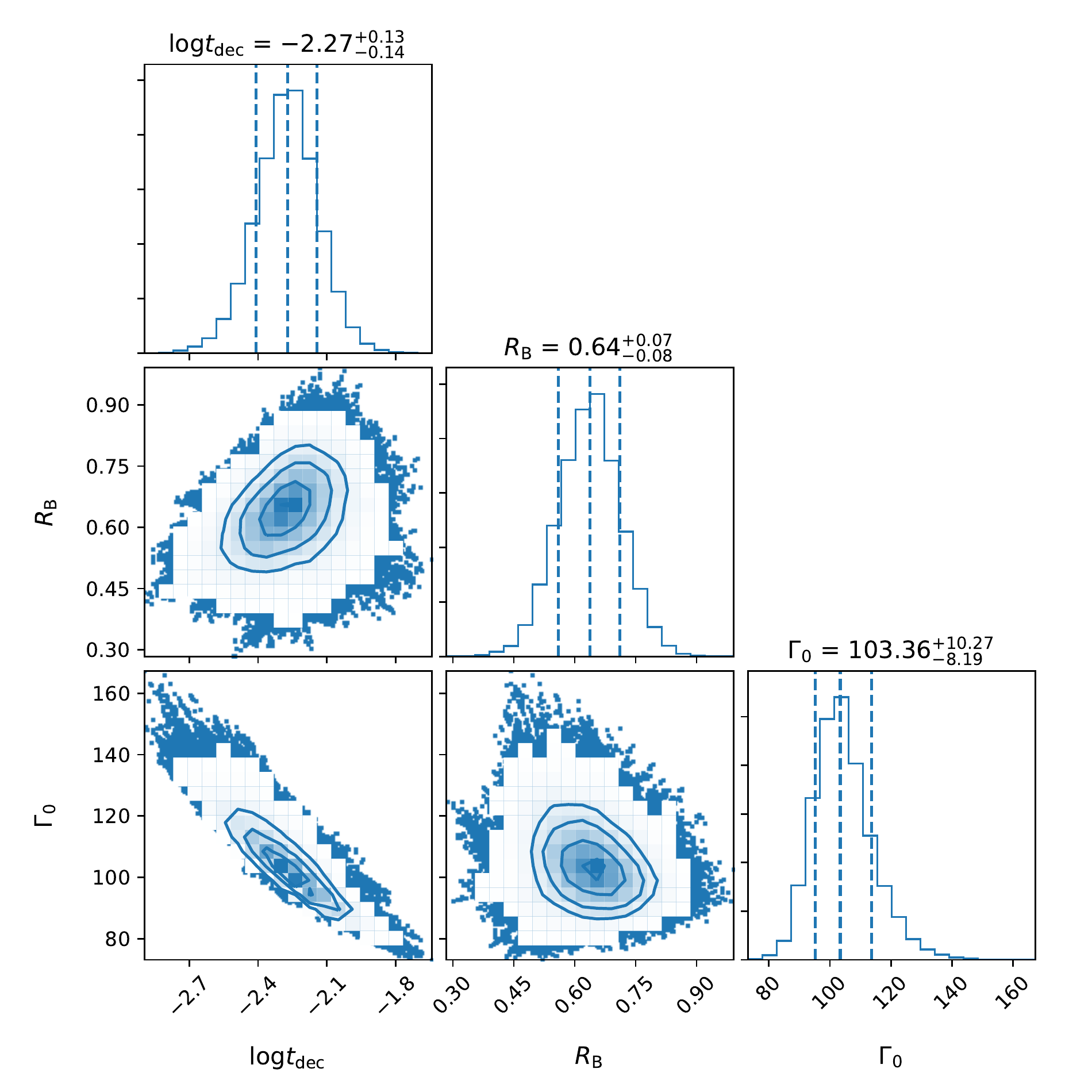}
  \caption{Correlations and marginalized posterior density functions for the deceleration time 
(\tdec) and relative magnetization ($\RB\equiv(\epsilon_{\rm B,r}/\epsilon_{\rm B,f})^{1/2}$) from 
the RS-FS consistency argument (equation \ref{eq:jetpropcalc}), and the jet initial Lorentz factor 
(\Gammajet), derived using energy balance (equation \ref{eq:Gammajet}).}
\label{fig:gamma}
\end{figure}

To infer $\Gammajet$ in this case, we turn to an energy balance argument \citep{zkm03}. At $\tdec$, 
the energy in the swept-up material ($E_{\rm sw}$) is similar to the rest energy of the ejecta 
(which is the same as the explosion kinetic energy, $E$),
\begin{align}
E_{\rm sw} = \Gammajet^2 M_{\rm sw} c^2 \approx E,
\label{eq:E_sw}
\end{align}
where the swept-up mass at radius, $R$,
\begin{align}
 M_{\rm sw} = \frac{4\pi A R^{3-k}}{3-k}
\label{eq:M_sw}
\end{align}
Combining equations \ref{eq:E_sw} and \ref{eq:M_sw}, the deceleration radius, 
\begin{align}
 R = \left[\frac{(3-k)E}{4\pi A\Gammajet^2 c^2}\right]^{1/(3-k)} = \Gammajet^{-2/(3-k)}\lsedov, 
\end{align}
where $\lsedov$ is the Sedov length, which depends only on the energy and density, parameters that 
are known from the FS modeling. 
For a point source moving along the line of sight, this radius corresponds to the deceleration time 
given by, $R = 2\Gammajet^2 c \tdec/(1+z)$,
which gives the initial Lorentz factor,
\begin{align}
 \Gammajet \approx 
\left[\frac{\lsedov}{c}\left(\frac{2\tdec}{1+z}\right)\right]^{\frac{3-k}{2(4-k)}}
\label{eq:Gammajet}
\end{align}
We note that this is a simplified treatment, and a more rigorous analysis requires integrating over 
the full fluid profile, and using the specific enthalpy of the shocked region while calculating 
the swept-up mass. However, for our purposes, this approximate expression will suffice to estimate 
\Gammajet. We compute $\Gammajet$ for all MCMC samples using equation~\ref{eq:Gammajet}, resulting 
in $\Gammajet = 103^{+10}_{-8}$ (Fig.~\ref{fig:gamma}).

\section{Discussion}
\label{text:discussion}
\subsection{Early optical rise}
\begin{figure}
  \centering
  \includegraphics[width=\columnwidth]{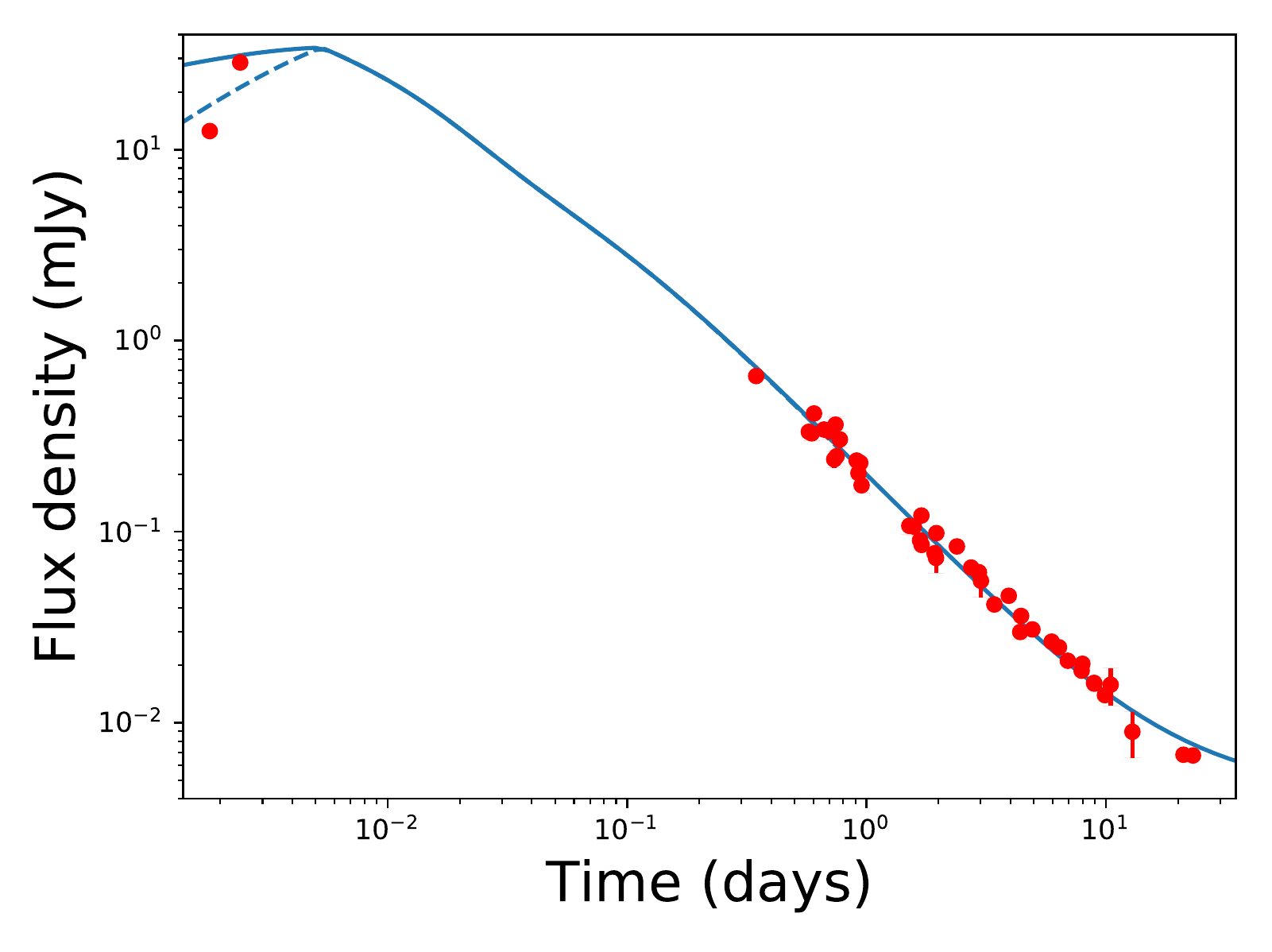}
  \caption{Optical light curve at $r^{\prime}/R$ and clear band light curve of GRB~181201, together 
with an FS model with energy injection at $\lesssim\tdec$ at the rate expected for a shell at 
a uniform Lorentz factor, $\Gammajet$, $E\propto t$ (solid) and at an enhanced rate of $E\propto 
t^{1.8}$ (dashed). The steep optical rise observed with MASTER suggests rapid energy injection into 
the FS during the deceleration stage (Section~\ref{text:optrise2}). }
\label{fig:enj}
\end{figure}

We now return to the optical rise observed at $\lesssim2.4\times10^{-3}$~days with MASTER  
(Fig.~\ref{fig:modellc_RS_splits}). These observations were performed at $\lesssim\tdec$ in our 
best-fit RS+FS model (Section~\ref{text:multimodel} and Table~\ref{tab:params}), consistent with 
the afterglow onset scenario. The spectral ordering at $\tdec$ is $\numr<\nuar<\nuopt\approx\nucr$ 
and $\nuaf<\nucf<\nuopt<\numf$. In this model, the RS light curve at $\lesssim\tdec$ is expected to 
rise at the rate $\alpha=(p-1)/2\approx0.5$, while the FS light curve should rise as $\alpha=1/2$ 
\citep{zwd05}. 
However, the observed rise of $\alpha\approx2.8$ is not consistent with either prediction. The 
proximity of $\nucr$ to $\nuopt$ at the deceleration time may suppress the RS flux. On the other 
hand, the FS model also over-predicts the observed flux. One possible solution is rapid injection 
of energy into the FS during this period. We find that an energy increase at the rate $E\propto 
t^{1.8}$ until $\tdec$ roughly accounts for the optical rise\footnote{The FS dynamics during shock 
crossing for a shell with a uniform \Gammajet\ is equivalent to energy injection at the rate 
$E\propto t$ \citep[Appendix B of][]{lbm+15}.} (Fig.~\ref{fig:enj}). Such an injection could arise, 
for instance, by the presence of energy at Lorentz factors $\lesssim\Gammajet$ owing to velocity
stratification in the ejecta \citep{sm00}, which has previously been inferred in other GRB 
afterglows \citep{lbm+15,lbm+18,lab+18}. An alternate scenario could be a pair-dominated 
optical flash, as suggested for GRB~130427A \citep{vhb14}. However, the paucity of data around the 
optical peak preclude a more detailed analysis. 

\label{text:optrise2}

\subsection{Intermediate millimeter excess}
\label{text:mmexcess}
\begin{figure}
  \centering
  \includegraphics[width=\columnwidth]{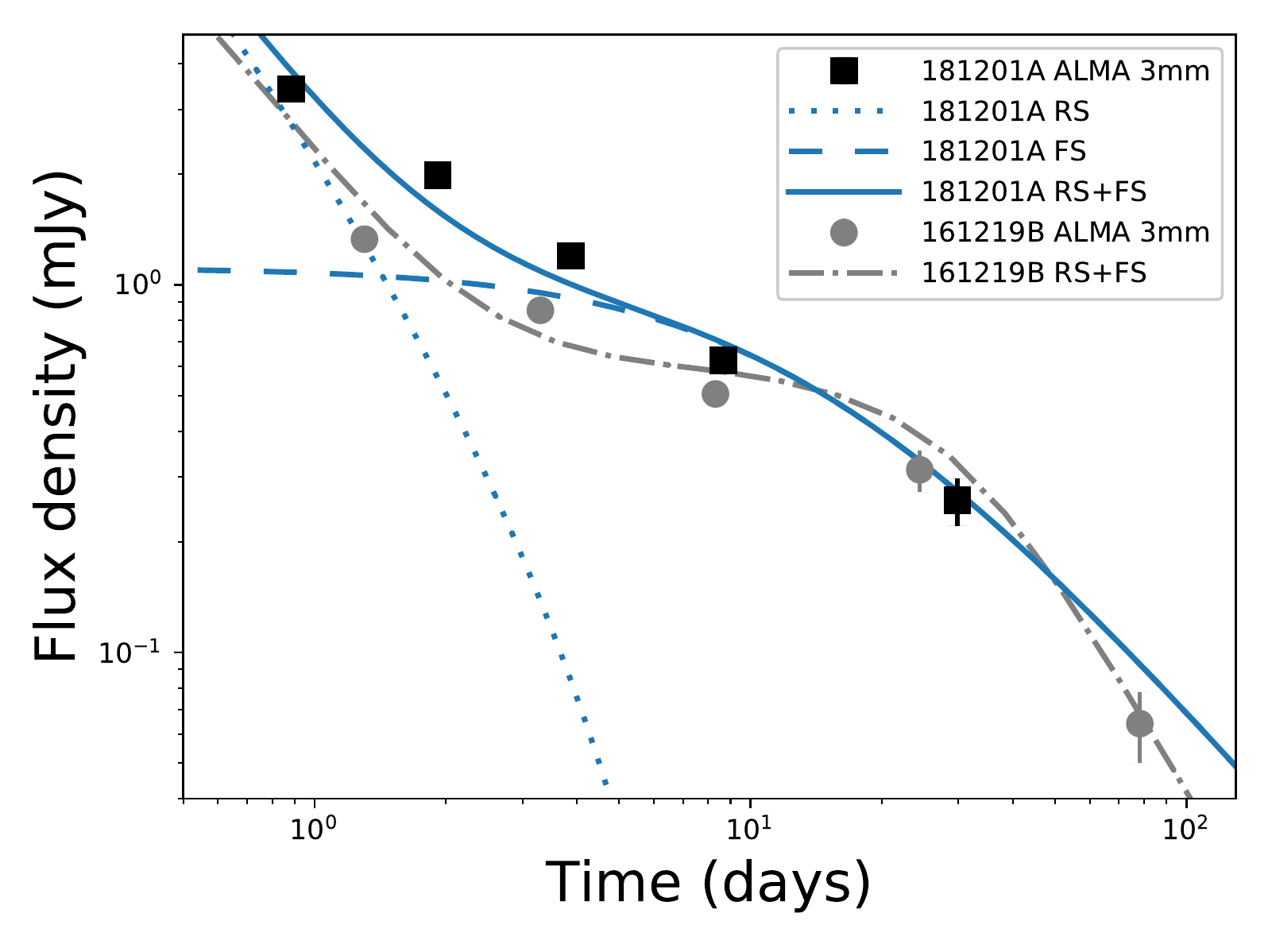}
  \caption{ALMA 3\,mm light curve of GRB 181201A (black squares) compared with that of 
GRB~161219B \citep[grey circles;][]{lab+18}, together with RS+FS models (lines). Both GRBs exhibit 
excess mm-band emission near the RS-FS transition, indicating additional physics not captured by the 
model. }
\label{fig:almacomparison}
\end{figure}

The 3\,mm model light curve under-predicts two of the ALMA detections at $\approx1.9$ and 
$\approx3.9$~days by $\approx20\%$ (Fig.~\ref{fig:almacomparison}). In particular, the 
observed light curve exhibits a shallower decline than afforded by the current model, where the 
transition between RS-dominant to FS-dominant at $\approx 1.5$~days results in an 
inflexion in the model light curve, which is not seen in the data. 
One possibility for a shallower RS decline rate is for the RS to be refreshed by 
slower ejecta catching up with the FS \citep{sm00}. This scenario requires a distribution 
of Lorentz factors in the jet, which has indeed been inferred in other events 
\citep{lbm+15,lbm+18,lab+18},
and remains feasible here.
We note that a similar 
discrepancy was present in the only other ALMA light curve of a GRB available at this date, that of 
GRB~161219B \citep{lab+18}. In that case, the model under-predicts the data at $\approx 3$~days, 
corresponding to a similar RS-FS transition. These similarities may indicate that the RS evolution 
is more complex than previously thought, or that the mm-band emission is not dominated by the RS at 
all, and is in fact a separate component. Further ALMA 3\,mm observations of GRBs, 
together with detailed multi-wavelength modeling, may help shed light on this issue. 

\subsection{Late X-ray excess}
\label{text:xrayexcess}
The observed 1\,keV light curve exhibits a late time shallowing (Fig.~\ref{fig:XRToptlc_pl}), and 
can therefore not be fit perfectly in our model (Fig.~\ref{fig:modellc_RS_splits}), where only 
steepenings are expected \citep[for instance, following a jet break; ][]{sph99}. As a result, our 
best-fit model under-predicts the X-ray flux density at 127~days. Above $\numf$, the afterglow 
light curves are indeed expected to become shallower at late times when the FS becomes 
non-relativistic at $\tnr$ \citep{wkf98}. One simple definition of $\tnr$ corresponds to the time 
when $\Gamma\beta\approx1$ (i.e., $\Gamma\approx\sqrt{2}$). For the best-fit model, these 
parameters correspond to $\tnr\approx220$~years, which cannot explain the X-ray excess at 
$\approx127$~days. 
Another possibility is that the late-time excess corresponds to a contaminating source within the 
XRT point spread function, which has been observed at least once before; however, this 
possibility is quite rare and unlikely. One way to test this is to obtain observations of the 
afterglow with \Chandra; however, at the time of writing, the target was within an extended 
Sun-constraint for \Chandra, and not visible.

We note that late-time excess X-ray flux has been uncovered in other events
\citep{fbm+14,mgl+15,lbc+18}. 
Suggested explanations have included 
dust echoes, inverse-Compton scattering, late-time radiation from spin-down of a magnetar central 
engine, and interaction of the FS with the shocked stellar wind at the edge of the stellar wind 
bubble \citep{sd07,ccf+08,lrz+08,mg11,gvem13,fbm+14,mgl+15}. For reference, the shock radius at 
127~days is $\approx30$~pc for our best-fit parameters, which is significantly larger than the mean 
distance between O stars in a typical massive stellar cluster \citep{fmm99}.

\subsection{Late radio excess}
\label{text:radioexcess}
The 1--2~GHz GMRT light curves appear to exhibit an unexpected brightening of the afterglow between 
30--48~days after the burst. The corresponding radio maps were dynamic range limited, with noise 
dominated by image deconvolution errors. Thus, it is possible that our reported uncertainties 
underestimate the true variation of the non-Gaussian image noise. In that scenario, the light curve 
may be consistent with being flat, and therefore host-dominated. We caution that while a similar 
brightening is evident in the GMRT $750$~MHz observations, they appear to be absent at $620$~MHz. 
The two bands were observed together as part of a broader frequency coverage with the upgraded GMRT 
(GWB) system, and no systematic errors are expected in the light curve produced from different parts 
of the observing band.

On the other hand, we note that late-time light curve flattenings or re-brightenings have been 
observed in several radio afterglows \citep{bck+03,pk04,alb+17,lbc+18,lab+18}.
If real, the re-brightening in this event might indicate (i) late-time energy injection due to 
slower ejecta shells\footnote{the Lorentz factor of the FS is $\Gamma\approx10$ at 40~days.} 
\citep{sm00,kbdg16}, which has been inferred in several instances \citep{lbm+15,lbm+18,lab+18}, or 
(ii) interaction of the FS with the wind termination shock, a scenario discussed above in the 
context of the X-ray light curve flattening (Section~\ref{text:xrayexcess}). Any contribution from 
the emerging supernova is expected to peak on $\gtrsim 10$~year time scales, and therefore an 
unlikely explanation for this excess \citep{bdg15}. The emergence of a counterjet, such as has 
been claimed in cm-band observations GRB~980703 at $\gtrsim500$~days after the burst 
\citep{ls04,phtp17} should occur at $\gtrsim \tnr\approx220$~yr for 
this event (Section~\ref{text:xrayexcess}), and is also disfavored. 

\subsection{Are we really seeing the reverse shock?}
\label{text:other_scenarios}
We note that our derived value of $g\approx2.7$, while higher than the theoretically expected 
upper limit for a wind environment \citep{zwd05}, is comparable to $g\approx2.3$ in GRB~140304A 
\citep{lbm+15}. Previous studies have found even higher values of $g\approx5$ in the case of the RS 
emission from GRB 130427A \citep{lbz+13,pcc+14}. Higher values of $g$ result in a more slowly 
evolving spectrum, and are preferred by the fit due to the relatively slowly evolving cm/mm emission 
observed in these events, including in the event studied here. Owing to this mismatch with the 
theoretical prediction for $g$, as well as some of the other issues described above, it is natural 
to ask whether the RS+FS model is the only one that can explain the observations. It is possible 
that the emission observed here and in similar events in the past, which has been ascribed to the 
RS, is in fact from a different emission mechanism. 

One possible alternate explanation is to invoke a two-component jet, one in which the optical and 
X-ray emission arises from a narrower, faster jet than that producing the cm-band observations 
\citep{bkp+03,pkg05,odpp+07,rks+08,hdpm+12}. In such a case, we expect the radio peak frequency and 
peak flux density to evolve according to the standard FS prescription 
as $\numax\propto t^{-3/2}$ and $\fnumax\propto t^{-1/2}$, respectively. 
From the SED at 3.9~days (Section~\ref{text:4dsed} and Fig.~\ref{fig:4dsed}), this would imply an 
$R$-band peak at $\approx2.2\times10^{-3}$~days with a flux density of $\approx60$~mJy. While this 
would over-predict the MASTER observations at $\approx2.4\times10^{-2}$~days, if the jet producing 
this emission component were still decelerating at this time, then the flux could be significantly 
lower. Furthermore, we find that this component would not contribute significantly to the remainder 
of the optical light curve. For instance, at the time of the earliest UVOT observation at 
$\approx0.18$~days, the predicted flux density of $\approx0.2$~mJy is below the observed flux 
density of $\approx0.6$~mJy. 
Whereas it is not possible for us to confirm or rule out two-component jet models, we note that 
decoupling the cm/mm emission from the optical/X-ray emission reduces the predictive power of 
such a model.

Another alternate scenario to explain the radio spectral bump at 3.9~days is the presence of a 
population of non-accelerated (``thermal'') electrons, which are not accelerated by the passage of 
the FS into a relativistic power-law distribution, but instead form a Maxwellian distribution at 
lower energies \citep{ew05}. However, a disjoint spectral component as observed here would likely 
require a very specific electron distribution, one in which the typical Lorentz factor of the 
thermal electrons is much lower than the minimum Lorentz factor of the shock-accelerated 
electrons, the so-called `cold electron model' \citep{rl17}. 

\begin{figure}
  \centering
  \includegraphics[width=\columnwidth]{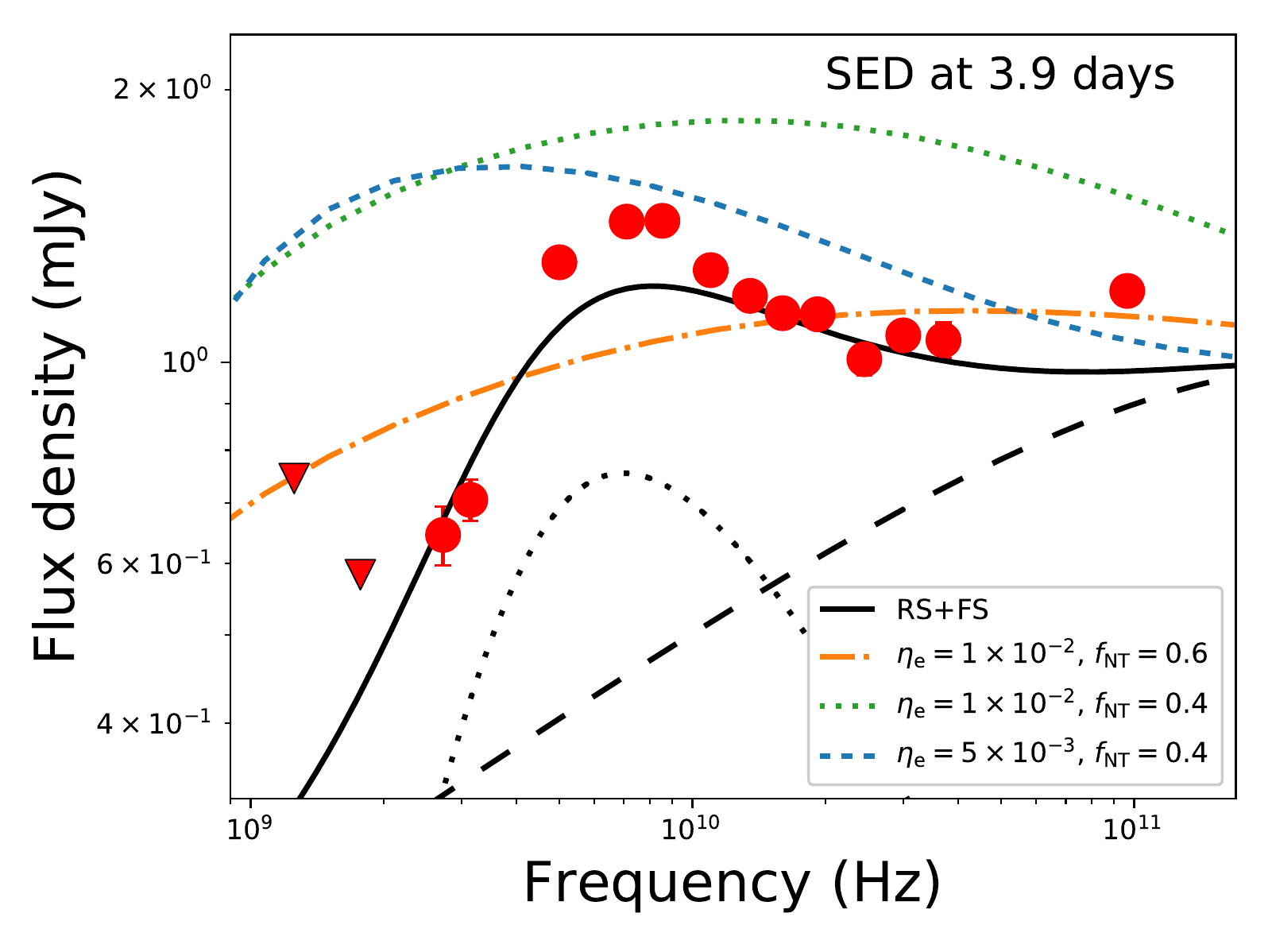}
  \caption{SED at $3.9$~days (data points), with our best-fit model (solid, black) decomposed 
into FS (dashed) and RS (dotted) components, compared with a model incorporating thermal electrons 
(orange dash-dot, blue short dashes, and green dotted). The thermal electron model cannot easily 
match the sharp spectral peak observed in 
the cm band.}
\label{fig:thermales}
\end{figure}

The emission signature from such a population is expected to be quite broad. We demonstrate this by 
setting the electron participation fraction, usually assumed to be unity, to $f_{\rm NT}\approx0.6$, 
and computing the resultant spectrum as described in \cite{rl17}. We use a cold electron model, with 
a ratio of electron temperature ($T_{\rm e}$) to gas temperature ($T_{\rm g}$) set at $\eta_{\rm e} 
\equiv T_{\rm e}/T_{\rm g}\approx10^{-2}$. The resultant spectrum exhibits a broad peak in the 
cm-band, and does not match the observations as well as the RS model (Fig.~\ref{fig:thermales}). 
While increasing $f_{\rm NT}$ increases the sharpness of the peak (relative to the underlying FS 
emission), doing so also increases the relative brightness of the thermal bump beyond the 
data, and so no combination of $f_{\rm NT}$ and $\eta_{\rm e}$ appears to match the cm-band 
data. A more detailed analysis requires a complete parameter search over $f_{\rm NT}$ and $\eta_{\rm 
e}$ that also includes the FS parameters. Such a study is currently not possible due to the 
computational intensiveness of the thermal electron calculation, and we defer it to future work.

\subsection{FS parameter degeneracies}
\label{text:IC}
Our modeling reveals degeneracies between the FS model parameters (Fig.~\ref{fig:corner}), which 
cannot be explained by the unknown value of $\nuaf$ alone. For instance, the correlation contours 
between $\Astar$ and $E_{\rm K,iso}$ lie parallel to the degenerate line $E_{\rm K,iso}\propto 
\Astar^{1.4}$. However, the expected relation between these parameters when $\nua$ is unknown is 
$E_{\rm K,iso}\propto \Astar^{-0.2}$ \citep{gs02,lbt+14}. We note that the Compton $Y$ parameter 
varies by about two orders of magnitude along the axis of the observed degeneracy, and thus the 
change in the physical parameters is compensated for by varying $Y$.  

Our prescription for calculating Compton $Y$ relies on a simplistic global approximation 
\citep{se01,lbm+15}. However, we note that in low density environments, the Klein-Nishina 
suppression limits the maximum value of $Y$ \citep{nas09}. We derive an expression for this maximum 
value of $Y$ corresponding to the Lorentz factor of cooling electrons in a wind environment by 
substituting equation~(18) from \cite{pk00} for the effective density as a function of observer time 
in equation~(64) from \cite{nas09} for the maximum value of $Y$ when KN corrections are included, to 
obtain, 
\begin{align}
 Y_{\rm max}(\gamma_{\rm c}) &= 10 e^{\frac{27(3.2-p)^2-17}{(p+2)(4-p)}}
                                   e^{\left(\frac{8-p}{2(p+2)}\ln{0.73}\right)}  \nonumber \\
                             &\times \epsilon_{\rm e,-1}^{\frac{5(p-1)}{p+2}}
                                E_{53}^{\frac{p-3}{p+2}}
                                \Astar^{\frac{8-p}{p+2}}
                                t_{\rm days}^{-\frac{(2p-1)}{(p+2)}}.
\end{align}
For our best-fit parameters (Table~\ref{tab:params}), this evaluates to 
$Y_{\rm max}(\gamma_{\rm c})\approx2 t_{\rm days}^{-0.8}$, which is lower than some of the values in 
the degenerate models discussed. Thus, a future, more accurate calculation incorporating
Klein-Nishina corrections may resolve this source of degeneracy. 

\subsection{Burst Energetics and radiative efficiency}
The data do not reveal a jet break \citep{rho99,sph99}, and hence the degree of collimation remains 
unknown for this burst. As of the time of writing, we are continuing VLA observations of this 
event, and expect that late-time observations may allow for beaming-independent calorimetry provided 
the contribution from the underlying host is negligible 
\citep{fwk00,bkp+03,bkf04,fsk+05,sb11,dcrgl12}. With that said, we note that our modeling does 
require some host contribution at 2.7 to 24.5~GHz at the $\approx50\,\mu$Jy level 
(Fig.~\ref{fig:modellc_RS_splits}), corresponding to a star formation rate of 
$\approx 2~M_{\odot}{\rm yr}^{-1}$ \citep{yc02,bck+03}; hence it is not clear whether such 
calorimetry will be feasible for this burst. For a jet break time of $\tjet\gtrsim163$~days, we can 
only set a lower limit to the jet half opening angle of $\thetajet\gtrsim6^{\circ}$, corresponding 
to a minimum beaming-corrected energy of $1.4\times10^{52}$~erg. 

The radiative efficiency, on the other hand, is independent of the beaming correction. For our 
best-fit model, we find $\etarad\approx0.35$. This value corresponds to a slightly lower efficiency 
than for other GRBs at similar $E_{\gamma,\rm iso}$ as derived from X-ray afterglow observations 
\citep{lz04}. However, the observed distribution of $\etarad$ for \Swift\ GRBs spans a broad range 
\citep{bkf03,nkg+06,zlp+07,bndp15}, and this value of $\etarad$ is unremarkable. 

\subsection{Low circum-burst density}
There is a strong observed correlation between multi-wavelength detections of RS emission with low 
density circum-burst environments, typically $\dens$, $\Astar\lesssim10^{-2}$ \citep{lab+18}. We 
note that the circum-burst density for our best-fit model, $\Astar\approx0.02$ is also similarly 
low. We have previously remarked on the possibility that this is an observational selection effect, 
since low density environments are more likely to result in a slow-cooling RS with long-lasting 
emission \citep{lbz+13,lab+16,lab+18,lag+19,kmk+15,alb+17}. In the case of GRB~161219B, we used 
constraints on the RS cooling frequency to determine a maximal particle density for the ISM 
environment under which RS emission was expected, and showed that the observed density satisfied 
the constraint in that case. We repeat that argument here, now for the wind medium.

Combining equation \ref{eq:consistency} with the expression for FS cooling frequency from 
\cite{gs02},
\begin{align}
 \nucf &= 4.40\times10^{10}(3.45-p)e^{0.45p}(1+z)^{-3/2} \nonumber\\
       &\times  \epsb^{-3/2}\Astar^{-2}E_{\rm K,iso,52}^{1/2}(1+Y_{\rm f})^{-2}\td^{1/2}
         {\rm Hz} \nonumber \\
       &\approx 10^{13} \Astar^{-2} {\rm Hz} \nonumber \\
       &\sim \RB^{3}\nucr 
\end{align}
for the best-fit parameters at $t\approx1$\,d yields,
\begin{align}
 \Astar &\approx \left(\frac{\nucr}{10^{13}\,{\rm Hz}}\right)^{-1/2}\RB^{-3/2}
\nonumber\\
 &\lesssim 0.24\RB^{-3/2}\pcc, 
\end{align}
which is indeed satisfied for GRB~181201A where $\Astar\approx2\times10^{-2}$ (corresponding to a  
mass loss rate, $\dot{M}\approx2\times10^{-7}~M_{\odot}{\rm yr}^{-1}$ for a wind speed of 
$10^3$~km\,s$^{-1}$). We note that a similar low wind parameter of $\Astar\approx3\times10^{-3}$ 
was inferred in the case of GRB~130427A, which also exhibited a strong RS signature 
\citep{lbz+13,pcc+14}. Our observation of RS emission in GRB~181201A therefore supports the 
hypothesis that RS emission is more easily detectable in low density environments, in both the ISM 
and wind scenarios. 

\subsection{Reverse shock Lorentz factor}
During our analysis above, we have assumed a non-relativistic reverse shock. We now check this 
assumption by calculating the Lorentz factor of the RS itself. During the shock crossing, the 
Lorentz factor of the shocked fluid in the cold ejecta frame is given by
\begin{align}
\bar{\gamma} &\approx \frac12\left(\frac{\Gamma_{\rm FS}/\sqrt{2}}{\Gammajet}
                      +\frac{\Gammajet}{\Gamma_{\rm FS}/\sqrt{2}}\right),
\end{align}
where $\Gamma_{\rm FS}$ is the FS Lorentz factor, assumed to be highly relativistic.
The RS Lorentz factor is then given by the shock jump condition \citep{bm76},
\begin{align}
\bar{\Gamma}_{\rm RS} = \frac{(\bar{\gamma}+1)\left[\hat{\varsigma}\left(\bar{\gamma}-1\right)+1\right]^2}{\hat{\varsigma}(2-\hat{\varsigma})(\bar{\gamma}-1)+2},
\end{align}
where we take the adiabatic index $\hat{\varsigma}=4/3$, appropriate for a relativistic fluid 
\citep{ks00}. We compute these quantities from our MCMC results, and find $\bar{\Gamma}_{\rm 
RS}=1.97^{+0.08}_{0.07}$, corresponding to a shock speed of $\bar{\beta}_{\rm RS}=0.74\pm0.02$ and 
4-velocity, $\bar{\Gamma}_{\rm RS}\bar{\beta}_{\rm RS}=1.7\pm0.1$. Thus the RS is at most mildly 
relativistic, and definitely not ultra-relativistic ($\bar{\Gamma}_{\rm RS}\bar{\beta}_{\rm 
RS}\gg1$). Whereas the expressions we have used for the RS evolution in the modeling are strictly 
valid only in the asymptotic regime of a Newtonian RS, they are expected to be approximately 
correct for cold shells where the post-shock evolution deviates from the Blandford-McKee solution. A 
more detailed analysis requires numerical calculation of the RS hydrodynamics and radiation 
simultaneously with the MCMC fitting, which is beyond the scope of this work. 

\subsection{Ejecta magnetization}
The relative magnetization of $\RB\approx0.6$ observed in this event is similar to $\RB\approx1$ for 
GRB~161219B \citep{lab+18} and $\RB\approx0.6$ for GRB~140304A \citep{lbm+18}. On the other hand, 
these values are all smaller than $1\lesssim\RB\lesssim5$ for GRB~130427A \citep{lbz+13}, 
$\RB\approx8$ for GRB~160509A \citep{lab+16}, $\RB\sim1$-100 for GRB~160509A \citep{alb+17}, 
and $\RB\approx1$--100 derived for GRBs with optical flashes \citep{gkg+08,gkm+09,jkck+14}.
Interpreting these values requires accounting for the variation in $\epsilon_{\rm B,f}$ between 
events, which we perform next.

The plasma magnetization in the ejecta, $\sigma$, defined as the ratio of the magnetic energy 
density to the rest mass energy density, corresponds to $\sigma\approx\epsilon_{\rm 
B,r}\bar{\Gamma}_{\rm RS}$. For the non-relativistic reverse shocks observed in all of these
cases, this implies $\sigma\approx\epsilon_{\rm B,r} = \RB^2\epsilon_{\rm B,f}$. 
A detailed statistical comparison of $\sigma$ between GRBs requires full posterior density 
functions for $\RB$ in each case, which are not currently available. However, we can use the 
typical parameters in each case for a rough comparison.
For our best-fit parameters, we have $\epsilon_{\rm B,r}\approx2.6\times10^{-3}$ for GRB~181201A. 
For GRB~160625B, the large feasible range of $\RB\sim1$--100 makes comparison less meaningful. For 
GRB~160509A, $\RB^2\epsilon_{\rm B,f}>1$; however, the RS in that case was long-lived and 
continuously refreshed, thus suggesting additional physics not captured by the simple calculation of 
$\sigma$.  For the other bursts with radio RS emission, we find $\epsilon_{\rm 
B,r}\approx6\times10^{-2}$ (161219B), 
$\epsilon_{\rm B,r}\approx2\times10^{-2}$ (140304A), $\epsilon_{\rm B,r}\approx 0.2$--1 (130427A). 
Thus for all GRBs with detected and well-constrained RS emission, $\sigma$ ranges between $10^{-3}$ 
and $\approx 1$, indicating that the ejecta are weakly magnetized. However, we caution that this 
outcome is largely built into the formalism for non-relativistic reverse shocks, for which 
$\sigma\approx\epsilon_{\rm B,r}$. The multiwavelength detection and characterization of a 
relativistic RS might enable us to explore a broader regime in $\sigma$.

In summary, the detection of the RS signature in this GRB argues against the scenario where the jet 
is Poynting flux dominated, since in that case, no RS emission is expected \citep{lcj17}. 
The mildly relativistic RS observed in this case argues against propagation in a highly magnetized, 
relativistic outflow, such as in a magnetar wind \citep{mgt+11}. This event brings the number of 
GRBs with radio RS detections to seven 
\citep{kfs+99,sp99a,lbz+13,pcc+14,lab+16,alb+17,lbm+18,lab+18,lag+19}.

\section{Conclusions}
We have presented multi-wavelength radio to X-ray observations of GRB~181201A spanning from 
$\approx150$~s to $\approx163$~days after the burst. We have modeled these observations in the 
standard afterglow synchrotron emission framework to derive the burst energetics and parameters of 
the explosion. The radio data we present reveal the presence of an additional component, which we 
interpret as emission from electrons heated by a mildly relativistic ($\bar{\Gamma}_{\rm 
RS}\approx2$) reverse shock that propagates through the ejecta in $\tdec\approx460$~s. We have 
performed detailed, self-consistent modeling of the RS and FS to derive the initial Lorentz factor, 
$\Gamma=103^{+10}_{-8}$ and magnetization, $\RB\equiv B_{\rm RS}/B_{\rm FS}=0.64^{+0.07}_{-0.08}$ of 
the jet that produced this gamma-ray burst. The low inferred ejecta magnetization disfavors a 
Poynting flux-dominated outflow. The derived low circum-burst density is consistent with the 
emerging requirement of low density environments for long-lasting, detectable RS emission to be 
produced.  The MCMC analysis performed here explores degeneracies between the RS and FS parameters 
for the first time, and can be used to derive the properties of the ejecta and surrounding 
environment. 
We have explored deviations of the data from the proposed model, including a steep early 
optical rise, an excess in the millimeter band on intermediate time scales, and a late-time excess 
in the X-ray and radio emission, and have placed these observations in the context of similar 
signatures observed in other events. Detailed multi-frequency observations and modeling of GRBs 
followed by a summative sample study may provide greater insight into the origin of these 
deviations.
Our results reveal additional degeneracies in the forward shock parameters due to an incomplete 
treatment of inverse Compton cooling; future work that incorporates the Klein-Nishina corrections 
may help alleviate this problem, allowing for even better constraints on the underlying physical 
processes and central engine responsible for GRBs and their afterglows.

\acknowledgements
We thank the anonymous referee for their helpful suggestions.
This paper makes use of the following ALMA data: ADS/JAO.ALMA\#2018.1.01454.T (PI: Laskar). ALMA is 
a partnership of ESO (representing its member states), NSF (USA) and NINS (Japan), together with 
NRC (Canada), NSC and ASIAA (Taiwan), and KASI (Republic of Korea), in cooperation with the Republic 
of Chile. The Joint ALMA Observatory is operated by ESO, AUI/NRAO and NAOJ. 
VLA observations for this study were obtained via project 18A-088 and 18B-407 (PI: Laskar). The 
National Radio Astronomy Observatory is a facility of the National Science Foundation operated 
under cooperative agreement by Associated Universities,  Inc. 
GMRT observations for this study were obtained via project 35\_065 (PI: Laskar). We thank the staff 
of the GMRT that made these observations possible. GMRT is run by the National Centre for Radio 
Astrophysics of the Tata Institute of Fundamental Research.
MASTER equipment is supported by the Development Program of Lomonosov MSU and by Moscow Union 
Optika. MASTER-SAAO is supported by National Research Foundation of South Africa. 
MASTER-Tunka equipment is supported by RFMSHE  grants 2019-05-592-0001-7293, 2019-05-595-0001-2496 .
We thank J.~Rupert for obtaining some of the MDM observations.
This work is based in part on observations obtained at the MDM Observatory, operated by Dartmouth 
College, Columbia University, Ohio State University, Ohio University, and the University of 
Michigan.
The Bath Astrophysics Group acknowledges support from the University of Bath and the Science and 
Technologies Facilities Council. 
KDA acknowledges support provided by NASA through the NASA Hubble Fellowship grant 
HST-HF2-51403.001 awarded by the Space Telescope Science Institute, which is operated by the 
Association of Universities for Research in Astronomy, Inc., for NASA, under contract NAS5-26555.
RBD acknowledges support from the National Science Foundation under Grant 1816694.
The Berger Time-Domain Group at Harvard is supported in part by NSF under grant AST-1714498 and by 
NASA under grant NNX15AE50G. 
RC thanks the Kavli Institute for Theoretical Physics for its hospitality while some of this work 
was completed. This research was supported in part by the National Science Foundation under Grant 
No. NSF PHY-1748958.
AG acknowledges the financial support from the Slovenian Research Agency (grants P1-0031, I0-0033, 
and J1-8136) and networking support by the COST Actions CA16104 GWverse and CA16214 PHAROS.
NJ is supported by a Jim and Hiroko Sherwin Studentship in Astrophysics.
RY is in part supported by Grants-in-Aid for KAKENHI grants No. 18H01232.
VL, EG, and NT acknowledge support by the Russian Foundation for Basic Research 17-52-80133.
DAHB acknowledges research support from the National Research Foundation of South Africa.
This work makes use of data supplied by the UK Swift Science Data Centre at the University of 
Leicester and of data obtained through the High Energy Astrophysics Science Archive Research Center 
On-line Service, provided by the NASA/Goddard Space Flight Center. 

\bibliographystyle{apj}
\bibliography{grb_alpha,gcn}
\end{document}